\documentclass[10pt,final,journal]{IEEEtran}

\pdfoutput=1
\usepackage[cmex10]{amsmath}
\usepackage{graphicx}
\usepackage{amsfonts}
\usepackage[scaled=0.92]{helvet}

\newcommand{\half}{\tfrac{1}{2}}
\newcommand{\var}{\mathop{\rm var} \nolimits}
\newcommand{\eig}{\mathop{\rm eig} \nolimits}
\newcommand{\Tr}{\mathop{\rm Tr} \nolimits}
\newcommand{\asymx}{\mathop{\sim}}
\newcommand{\asym}[1]{\mathrel{\asymx_{#1}}}
\newcommand{\diag}{\mathop{\rm diag} \nolimits}
\newcommand{\sgn}{\mathop{\rm sgn} \nolimits}

\newcommand{\Imz}{\mathop{\rm Im} \nolimits}

\newcommand{\MGF}{\mathcal{M}}
\newcommand{\CGF}{\mathcal{K}}
\newcommand{\mtx}[1]{\mathsf{#1}}

\begin{document}
\title{Correlated Non-Coherent Radar Detection for Gamma-Fluctuating Targets
       in Compound Clutter}

\author{Josef~Zuk\\
Defence Science and Technology Group, Australia%
\thanks{E-mail: josef.zuk@dst.defence.gov.au}}

\maketitle

\begin{abstract}
This work studies the problem of radar detection
of correlated gamma-fluctuating targets
in the presence of clutter described by
compound models with correlated speckle.
If the correlation is not accounted for in a radar model, the required
signal-to-interference ratio for a given probability of detection will
be incorrect, resulting in over-estimated performance.
Although more generally applicable, the is focus on airborne
maritime radar systems.
Hence K-distributed sea clutter is used as the main example.
Detection via square-law non-coherent pulse integration
is formulated in a way that accommodates arbitrary partial correlation
for both target radar cross-section (RCS) and clutter speckle.
The obstacle to including this degree of generality in previous
work was the fact that Swerling's original characterization of the standard RCS
fluctuation classes as gamma distributions for the power is not sufficient for
the inclusion of both correlation sources
({\it i.e.}\ target and clutter speckle)
for gamma-fluctuating targets. An
extension of the model is required at the quadrature component
({\it i.e.}\ voltage) level, as phase relationships can no longer be neglected.
This is addressed in the present work, which not only postulates an extended model,
but also demonstrates how to efficiently compute it, with and without
a number of simplifying approximation schemes within the framework
of the saddle-point technique.
\end{abstract}


\section{Introduction}
\label{intro}

\IEEEPARstart{P}{arametric} modelling is a useful tool for predicting the
detection performance of a radar system given a statistical description of the
environment. Statistical models of sea clutter are commonly employed in the
parametric modelling of detection by maritime surveillance radar \cite{GC:WTW13,GC:Rosen17}.
These are often compound models that represent the fast varying speckle component
of the clutter with a local Gaussian process which is modulated by a slowly varying
texture component. If the radar employs frequency agility, it may be possible to
treat the speckle as completely decorrelated from pulse to pulse, however if the
radar uses a fixed frequency or the agile frequencies are separated by less than the
pulse bandwidth, the speckle correlation must be taken into account.

In this paper, the problem of radar detection using non-coherent pulse integration is
considered, where the fluctuations of target RCS are typically
modelled in terms of the standard Swerling classes.
Varying degrees of temporal correlation of target RCS and
clutter speckle fluctuations are relevant considerations for performance modelling of
airborne maritime surveillance systems.
Therefore, we formulate a modelling scheme that accommodates the simultaneous inclusion
of arbitrary partial correlation of both a gamma-fluctuating target
RCS and the speckle
component of a compound clutter model. In line with the maritime focus of the paper,
K-distributed sea clutter is used as the main example.

In the case where the only source of interference is (uncorrelated) thermal noise,
general exact calculations for detection probability resulting from non-coherent pulse
integration and partially correlated Rayleigh fluctuating target RCS
were given by Kanter \cite{GC:Kanter86},
and later for general gamma-fluctuating targets by Weiner \cite{GC:Weiner88}.
Various simplified special cases had been previously considered, and a review of these
can be found in \cite{GC:Buterbaugh92}.
A discussion of Kanter's model in which both the target RCS and clutter
returns are present and partially correlated was given in \cite{GC:Hou88}.
Subsequently, efficient approximations to the exact but
numerically unstable results of Kanter and Weiner have been
developed, based on
effective number-of-looks concepts \cite{GC:WTW13,GC:Rosen17},
and saddle-point techniques \cite{GC:Helstrom92,GC:Bocquet18,GC:Zuk19}.
The problem of correlated gamma-fluctuating targets in uncorrelated clutter
has also recently been studied in \cite{GC:Yang18}.

In previous work on the saddle-point approach to the problem \cite{GC:Zuk19},
attention was restricted to uncorrelated clutter speckle because
the full problem of a correlated gamma-fluctuating target in correlated
clutter speckle requires considerations beyond those of computational
methodology (saddle-point or otherwise); in particular, an extension of the
underlying model must first be established.
The classic Swerling models of target RCS fluctuation are specified in terms of the power
distributions of the target returns \cite{GC:Swerling57}. While this suffices generally for
exponentially fluctuating targets,
when the target power fluctuations are gamma distributed
and when {\em both} the target and clutter returns are correlated, the distribution of total
returned power depends also on the relative phases of target quadrature components \cite{GC:Zuk18}
due to a breaking of orthogonal symmetry.
A similar issue for clutter is discussed in \cite{GC:Conte87}, where it is
pointed out that the relative phase between quadrature components is correlated
even in the Gaussian problem unless it comprises entirely white noise.
A general first-principles model that specifies the joint probability distribution
of the quadrature components in a way
that reduces correctly to all known special cases
and satisfies all expectations has not previously appeared in the literature. A solution to this
problem, along with the demonstration of the applicability of the saddle-point technique, is
the subject of the present paper.
While attention here is focussed on pulse-to-pulse integration, the concepts developed
are also applicable to systems employing scan-to-scan integration \cite{GC:Rosen14}.

The rest of the paper is organized as follows:
Section~\ref{ProbForm} introduces the problem and the general method of solution using the
inverse Laplace transform of the moment generating function.
Section~\ref{CorClutter} describes the representation of the moment generating function for
correlated clutter speckle.
Section~\ref{FirstPrinc} develops the exact physical solution to the problem, designated as a
`first principles' model, which is not solvable by analytic means.
Section~\ref{EffModel} derives an effective method to evaluate the first-principles model.
The resulting effective model serves as a proxy for the first-principles model for
computational purposes.
A number of approximations are then proposed to further reduce the computational load while
maintaining robust performance for all realistic conditions.
In Section~\ref{mcsim}, Monte Carlo simulation is employed to demonstrate the high degree of
accuracy of the effective method in approximating the first-principles model,
in all physically realistic situations.
It should be noted that
the effective model does not stand on its own:
If the first-principles-model were not developed, then
there would be no basis upon which to judge whether the
effective model is a valid representation of the problem.
Section~\ref{Results} then summarizes the key results in the paper.

\section{Problem Formulation}
\label{ProbForm}
To be specific, we consider square-law detection followed by
$M$-pulse non-coherent integration, in the presence of multiple partially
correlated signal/noise power sources.
The  test statistic that determines the detection threshold
is the summed returned power random variable (RV)
\begin{equation}
Z = \half\sum_{\eta=\text{i},\text{q}}
     \Bigl|\sum_\alpha \mathbf{X}_{\eta\alpha} \Bigr|^2 \,,
\end{equation}
where $\mathbf{X}_{\eta\alpha}$ is an $M$-dimensional vector with elements
$X_{\eta\alpha m}$,
\mbox{$m = 1,2,\ldots,M$}.
Vector RVs are indicated by a bold typeface. For
\mbox{$\eta = \text{i},\text{q}$},
the $X_{\eta\alpha m}$ are respectively the in-phase and quadrature components
on the $m$-th integrated pulse for signal power source $\alpha$.
The source index $\alpha$ will typically range over thermal noise (n), target return (s)
and various types of clutter (c).
The target RCS fluctuation class indexed by
\mbox{$\kappa = 1,2,\ldots$},
refers to all models that interpolate (in degree of correlation)
between the  Swerling-($2\kappa-1$)
and Swerling-($2\kappa$) limiting cases.
Thus, the
\mbox{$\kappa = 1$}
class interpolates between the fully correlated Swerling 1 and
fully uncorrelated Swerling 2 models, while the
\mbox{$\kappa = 2$}
class interpolates between the fully correlated
Swerling 3 and fully uncorrelated Swerling 4 models.
A steady target RCS (Swerling 0) is obtained asymptotically as
\mbox{$\kappa\rightarrow\infty$}.

The probability density function (PDF)
\mbox{$P_Z(z)$}
of the RV $Z$ is the inverse Laplace transform of the
moment generating function (MGF)
\begin{equation}
\MGF_Z(s) \; \equiv \; \left\langle e^{-sZ}\right\rangle_{Z} \; = \;
     \left[\left\langle e^{-\frac{s}{2}\left(\sum_\alpha\mathbf{X}_\alpha\right)^2}
     \right\rangle_{\mathbf{X}_\alpha}\right]^2 \,,
\label{LZ}
\end{equation}
where $\mathbf{X}_\alpha$,
representing a {\em single} quadrature component,
is distributed identically with both $\mathbf{X}_{\text{i}\alpha}$ and
$\mathbf{X}_{\text{q}\alpha}$.
The square on the RHS of the equation above accounts for the presence of
two quadrature components.
The notation $\langle\cdot\rangle_Z$ denotes the
expectation with respect to the distribution of the random variable $Z$.
If we initially restrict ourselves to the
\mbox{$\kappa = 1$}
target RCS fluctuation class then,
for each power source $\alpha$,
the joint PDF of the vector RV $\mathbf{X}_\alpha$
is a mean-zero correlated Gaussian with variance
$\sigma^2_\alpha$
representing the returned power associated with power source $\alpha$,
and correlation matrix $\mtx{C}_\alpha$,
It should be noted that the matrices $\mtx{C}_\alpha$ are taken to be
correlation matrices, not covariance matrices.
Therefore, it is assumed that
\mbox{$\Tr\{\mtx{C}_\alpha\} = M$}.
For the thermal noise component, the correlation matrix is the identity:
\mbox{$\mtx{C}_{\text{n}} = \mathbb{I}$}.

It is convenient to introduce the notation
\mbox{$Z_{\text{avg}}\equiv Z/M$}
for the pulse-averaged returned power, and
\mbox{$Z'_{\text{avg}} \equiv  Z_{\text{avg}}/\sigma^2_{\text{I}}$}
for the average power normalized by the mean total interference
\mbox{$\sigma^2_{\text{I}} \equiv \sigma^2_{\text{n}} + \sigma^2_{\text{c}}$}.
Also,
\mbox{$S \equiv \sigma^2_{\text{s}}/\sigma^2_{\text{I}}$}
will denote the signal-to-interference ratio (SIR),
and
\mbox{$q \equiv \sigma^2_{\text{c}}/\sigma^2_{\text{I}}$}
(\mbox{$0 \le q \le 1$})
is the clutter-to-interference ratio.

We shall work with the
Gauss-Markov correlation model \cite{GC:Kanter86}, for which
the correlation matrices have the symmetric Toeplitz form:
\begin{equation}
[\mtx{C}_{\text{s}}]_{mn} = \rho_{\text{s}}^{|m-n|} \,, \quad
[\mtx{C}_{\text{c}}]_{mn} = \rho_{\text{c}}^{|m-n|} \,,
\label{GMC}
\end{equation}
where $\rho_{\text{s}}$, $\rho_{\text{c}}$ are the target and clutter
correlation coefficients, respectively, such that
\mbox{$0 \le \rho_{\text{s}}, \rho_{\text{c}} \le 1$}.
Nevertheless, the methodology that we develop works equally
with any other Toeplitz correlation structure.

By considering the various known special cases, such as correlated
exponentially fluctuating target power
(\mbox{$\kappa = 1$})
with correlated clutter speckle \cite{GC:Bocquet18,GC:Zuk18}, and
correlated gamma-fluctuating target power
(\mbox{$\kappa > 1$})
with uncorrelated clutter speckle \cite{GC:Zuk19},
one can infer that the mean and variance of the total returned
power in the general case should be given, respectively, by
\begin{equation}
\left\langle Z_{\text{avg}}/\sigma^2_{\text{I}}\right\rangle = 1 + S \,,
\label{MnVar1}
\end{equation}
and
\begin{equation}
\var\left(Z_{\text{avg}}/\sigma^2_{\text{I}}\right)
     = \frac{1-q^2}{M} + \zeta{\cdot}\frac{q^2}{L} + \frac{S^2}{\kappa B}
     + 2S{\cdot}\left(\frac{1-q}{M} + \frac{q}{N}\right) \,,
\label{MnVar2}
\end{equation}
where
\begin{equation}
L \equiv M^2/\Tr\{\mtx{C}^2_{\text{c}}\}, \, B \equiv M^2/\Tr\{\mtx{C}^2_{\text{s}}\}, \,
     N \equiv M^2/\Tr\{\mtx{C}_{\text{c}}\mtx{C}_{\text{s}}\} \,.
\label{MnVar3}
\end{equation}
The quantity $\zeta$ represents the effect of compound clutter texture,
and is given by
\begin{equation}
\zeta \, = \, 1 + (L+1)\var(U) \, = \, 1 + (L+1)/\nu \,,
\label{MnVar4}
\end{equation}
with the second equality holding for K-distributed compound clutter,
where the texture RV $U$ is described by a unit-mean gamma distribution
with shape parameter
equal to the clutter texture shape parameter $\nu$, in which case
\mbox{$\var(U) = 1/\nu$}.
Standard Rayleigh clutter corresponds to
\mbox{$\zeta = 1$},
obtained in the limit
\mbox{$\nu\to\infty$}.
It should be noted that
the considerations in this work remain valid for any compound clutter model.
In the general case, $U$ is the random variable that realizes the clutter
texture distribution in any compound clutter model.

The simplest way to understand (\ref{MnVar4}) is to set
\mbox{$S = 0$}
and
\mbox{$q = 1$}.
Then
\mbox{$Z_{\text{avg}}/\sigma^2_{\text{I}} = UV$},
where the clutter speckle RV $V$ is given by the correlated sum
of $M$ unit-mean exponential RVs (and scaled by $M$).
Since the second moment of the compound RV factorizes into
texture and speckle components, we have
\mbox{$\var\left(Z_{\text{avg}}/\sigma^2_{\text{I}}\right) = m_2(U) m_2(V) - 1$},
given that
$\langle U\rangle = \langle V\rangle = 1$.
The result follows upon observing that
\mbox{$\var(V) = 1/L$} \cite[Eq.~65]{GC:Kanter86}.

The challenge is to derive a general model that reproduces this mean and variance.
It is worth mentioning at this point that the related problem of including partial
correlation of clutter speckle in cases where a non-Rayleigh overall clutter
amplitude is being modelled was successfully addressed by the introduction of the
spherically invariant process (SIRP) \cite{GC:Conte87},
of which the compound clutter paradigm is a special case.
However, this approach is not applicable to the correlation of non-Rayleigh
target amplitudes because introducing a SIRP leads to an increase in power variance,
as can be seen for the clutter contribution in (\ref{MnVar2}) and (\ref{MnVar4})
by comparing a finite shape parameter $\nu$ with
\mbox{$\nu\to\infty$};
On the other hand, moving the RCS fluctuation class index $\kappa$ to
values greater than unity will necessarily decrease the variance.

Compound clutter is incorporated in the usual way \cite{GC:Bocquet18,GC:Zuk18}
by promoting the mean clutter power to a random variable according to the mapping
\mbox{$\sigma_{\text{c}}^2 \mapsto \bar{\sigma}_{\text{c}}^2{\cdot}U$}
where the RV $U$ generates the clutter texture distribution with unit mean.
The overall mean interference power becomes
\begin{equation}
\sigma_{\text{I}}^2 \; = \; \sigma_{\text{n}}^2 +
     \left\langle\sigma_{\text{c}}^2(U)\right\rangle_U
     \; = \; \sigma_{\text{n}}^2 + \bar{\sigma}_{\text{c}}^2 \,.
\end{equation}
The texture expectation is best left to the end, after recovering the survival
function from the MGF. Thus, if
\mbox{$\bar{F}_{\text{cc}}(v;S)$}
denotes the survival function for the total returned power $v$
in the presence of both a target and compound clutter,
and we also use the notation
\mbox{$\bar{F}_{\text{spk}}(v;S,q)$},
where
\mbox{$q \equiv \bar{\sigma}_{\text{c}}^2/\sigma_{\text{I}}^2$}
is the clutter-to-interference ratio,
to denote the survival function incorporating clutter speckle
but without the texture, then
\mbox{$\bar{F}_{\text{cc}}(v;S)$}
is given by
\begin{align}
\begin{aligned}
\bar{F}_{\text{cc}}(v; S) &= \left\langle
     \bar{F}_{\text{spk}}(v; S, qU)\right\rangle_U \\
&= \int_0^\infty du\, f_\nu(u)
     \bar{F}_{\text{spk}}(v; S, qu) \,,
\end{aligned}
\label{Fspk}
\end{align}
where
\mbox{$f_\nu(u)$}
denotes the clutter texture PDF. For K-distributed clutter, this is a unit-mean
gamma distribution with shape parameter $\nu$ \cite{GC:WTW13}.
The texture integration is most efficiently performed by means of
Gaussian quadrature.

The probability of detection of a target on a single scan $P_{\text{D}}(S)$
when its SIR is known to be $S$, is given by the value of the survival
function (also known as the complementary cumulative distribution function)
for the total returned (signal plus interference) power at the
selected detection threshold
\mbox{$v = v_{\text{b}}$}.
The threshold is determined as the point at which the survival function for
the total returned interference power assumes the value that corresponds to
a pre-determined desired probability of false alarm $P_{\text{FA}}$.
Thus, we eliminate $v_{\text{b}}$ in the following pair of equations:
\begin{equation}
P_{\text{D}}(S) = \bar{F}_{\text{cc}}(v_{\text{b}};S) \,, \quad
     P_{\text{FA}} = \bar{F}_{\text{cc}}(v_{\text{b}};0) \,.
\label{PdPfa}
\end{equation}

In the saddle-point method, just as for the PDF,
one computes the survival function in terms of
the inverse Laplace transform of the MGF for its underlying distribution,
by integrating along a suitably constructed contour in the complex
plane \cite{GC:Helstrom92}.

Previous work \cite{GC:Zuk19} applied saddle-point techniques to the problem of
correlated targets in uncorrelated clutter speckle, using an existing model
for the problem. The present work, in part, also applies saddle-point
techniques, but to a model that must first be developed, as there is no extant
model that addresses the problem of correlated gamma-fluctuating target in
correlated clutter speckle.

\section{Correlated Clutter}
\label{CorClutter}
Weiner's model \cite{GC:Weiner88} is applicable to correlated gamma-fluctuating
targets embedded in uncorrelated clutter speckle.
In this section, we generalize this by establishing the representation
of the MGF that incorporates speckle correlation,
and discuss the complications encountered when trying to evaluate it
according to the usual specification of the Swerling RCS models.

We assume that there are three sources of returned power
\mbox{$\alpha = \text{n},\text{c}, \text{s}$},
corresponding to thermal noise, surface clutter and target RCS, respectively,
and that the correlated gamma-fluctuating target is embedded in
correlated clutter (belonging to the
\mbox{$\kappa = 1$}
speckle fluctuation class).
The $M$-dimensional vector RV $\mathbf{X}_\alpha$,
composed of one element for each of the
\mbox{$m = 1,2,\ldots,M$}
pulses, will denote the in-phase quadrature component for the associated power sources.

The MGF
\mbox{$\MGF_Z(s)$}
of the returned-power RV $Z$ is given by \cite{GC:Zuk18}
\begin{align}
\begin{aligned}
\sqrt{\MGF_Z(s)} &=  \left\langle e^{-\frac{s}{2}\left(\sum_\alpha\mathbf{X}_\alpha\right)^2}
     \right\rangle_{\mathbf{X}_\alpha} \\
&= \int_{-\infty}^{+\infty} \frac{d^M u}{\left(2\pi s\right)^{M/2}}\,
     e^{-\mathbf{u}\cdot\mathbf{u}/(2s)}
     \prod_{\alpha=\text{n},\text{c},\text{s}}\left\langle e^{-i\mathbf{u}\cdot\mathbf{X}_\alpha}
     \right\rangle_{\mathbf{X}_\alpha} \\
&= \int_{-\infty}^{+\infty} \frac{d^M u}{\left(2\pi s\right)^{M/2}}\,
     e^{-\mathbf{u}^{\sf T}\mtx{Q}(s)\mathbf{u}/2}
     \left\langle e^{-i\mathbf{u}\cdot\mathbf{X}_{\text{s}}}\right\rangle_{\mathbf{X}_{\text{s}}} \,,
\end{aligned}
\label{XQ}
\end{align}
where we have split off the non-Gaussian target component
\mbox{$\mathbf{X}_{\text{s}}$},
evaluated the expectations over the clutter and noise components,
and set
\begin{equation}
\mtx{Q}(s) \equiv (1/s + \sigma^2_{\text{n}})\mathbb{I} + \sigma^2_{\text{c}}\mtx{C}_{\text{c}} \,.
\label{Qmatrix}
\end{equation}
The square-root on the LHS of the equation arises from the square on the RHS in (\ref{LZ}).

Now, since $\mtx{Q}$ is a symmetric matrix, we may write
\mbox{$\mtx{Q} = \mtx{R}^{\sf T}\mtx{\Lambda}\mtx{R}$},
where
\mbox{$\mtx{\Lambda} = \diag(\lambda_1,\ldots,\lambda_M)$}
and
\mbox{$\mtx{R}^{\sf T}\mtx{R} = \mathbb{I}$}.
Then we have, for any vector $\mathbf{J}$, the representation
\begin{align}
\begin{aligned}
&\int_{-\infty}^{+\infty} \frac{d^M u}{\left(2\pi\right)^{M/2}}\,
     e^{-\mathbf{u}^{\sf T}\mtx{Q}\mathbf{u}/2}{\cdot}e^{-i\mathbf{u}^{\sf T}\mathbf{J}} \\
&= \int_{-\infty}^{+\infty} \frac{d^M u}{\left(2\pi\right)^{M/2}}\,
     e^{-\mathbf{u}^{\sf T}\mtx{\Lambda}\mathbf{u}/2}{\cdot}e^{-i\mathbf{u}^{\sf T}\mtx{R}\mathbf{J}} \\
&= \frac{1}{\sqrt{\det\mtx{Q}}}\exp\left[-\sum_{m=1}^M \left(RJ\right)^2_m/(2\lambda_m)\right] \,,
\end{aligned}
\label{JQ}
\end{align}
where the first equality is due to a rotational change of integration variables
\mbox{$\mathbf{u}' = \mtx{R}\mathbf{u}$},
and we note that
\begin{align}
\begin{aligned}
\sum_{m=1}^M \left(\mtx{R}J\right)^2_m/\lambda_m &= \sum_{m,a,b = 1}^M J_a\left(R^{\sf T}_{am}
     \lambda^{-1}_m \mtx{R}_{mb}\right)J_b \\
&= \mathbf{J}^{\sf T}\mtx{Q}^{-1}\mathbf{J} \,,
\end{aligned}
\end{align}
which leads to the result
\begin{equation}
\sqrt{\MGF_Z(s)} = \frac{1}{s^{M/2}\sqrt{\det\mtx{Q}(s)}}
     \left\langle e^{-\frac{1}{2}\mathbf{X}_{\text{s}}^{\sf T}\mtx{Q}^{-1}(s)\mathbf{X}_{\text{s}}}
     \right\rangle_{\mathbf{X}_{\text{s}}} \,,
\label{XQX}
\end{equation}
upon making the identification
\mbox{$\mathbf{J} \equiv \mathbf{X}_{\text{s}}$}.

It is convenient to introduce a scaled target RV $\hat{\mathbf{X}}_{\text{s}}$ according to
\mbox{$\mathbf{X}_{\text{s}} = \sigma_{\text{s}}\hat{\mathbf{X}}_{\text{s}}$},
so that each power component $\hat{X}^2_{\text{s}m}$ has unit mean.
For RCS fluctuation class $\kappa$, each $\hat{X}^2_{\text{s}m}$ has a marginal distribution
given by the unit-mean gamma distribution with shape parameter $\kappa/2$, {\it i.e.}
\mbox{$\hat{X}^2_{\text{s}m} \sim \Gamma(\kappa/2, 2/\kappa)$}
for all
\mbox{$m = 1,2,\ldots,M$} \cite{GC:Swerling57}.

For uncorrelated clutter speckle, $\mtx{Q}$ becomes proportional to the unit matrix
\mbox{$\mtx{Q} = \xi^2\mathbb{I}$},
with
\begin{equation}
\xi^2 \; \equiv \; 1/s + \sigma^2_{\text{n}} +  \sigma^2_{\text{c}}
     \; = \; 1/s + \sigma^2_{\text{I}}\,.
\end{equation}
This yields the representation
\begin{equation}
\sqrt{\MGF_Z(s)} = \frac{1}{(s\xi^2)^{M/2}}
     \left\langle\exp\left(-\frac{\sigma^2_{\text{s}}}{2\xi^2}\hat{\mathbf{X}}^2_{\text{s}}\right)
     \right\rangle_{\hat{\mathbf{X}}_{\text{s}}} \,,
\end{equation}
which reproduces the problem considered in the previous work \cite{GC:Zuk19}
when combined with the expression \cite{GC:Weiner88}
\begin{equation}
\left\langle\exp\left(-\frac{\sigma^2_{\text{s}}}{2\xi^2}\hat{\mathbf{X}}^2_{\text{s}}\right)
     \right\rangle_{\hat{\mathbf{X}}_{\text{s}}} =
     \left[\det\left(1 + \frac{\sigma^2_{\text{s}}}{\kappa\xi^2}C_{\text{s}}
     \right)\right]^{-\kappa/2} \,.
\label{KapExp}
\end{equation}

The higher-order Swerling models
(\mbox{$\kappa > 1$})
are defined by giving the common distribution of the squares of the quadrature
components $\hat{X}_{\text{s}m}^2$ \cite{GC:Swerling57},
but the signs are left unspecified. In the presence of partial correlation, however,
the signs of the target quadrature components do not drop out and, therefore,
the specification is incomplete.
Equivalently, one may note that while
\mbox{$\hat{\mathbf{X}}^2_{\text{s}}$}
is rotationally invariant, the quadratic form
\mbox{$\hat{\mathbf{X}}_{\text{s}}^{\sf T}\mtx{Q}^{-1}\hat{\mathbf{X}}_{\text{s}}$}
is not.

\section{A First-Principles Model}
\label{FirstPrinc}
In this section, we review the derivation of Weiner's model for gamma-fluctuating
targets in uncorrelated interference, and use it to motivate a generalization to
accommodate the presence of clutter with a arbitrarily correlated speckle component.
The aim is to derive a first-principles model, {\it i.e.}\
one whose degrees of freedom correspond to
basic observable physical quantities.
In the present context, these are provided by the electromagnetic field
impinging upon the antenna aperture or, equivalently,
by the voltage variations induced because of it.
Due to the stochastic nature of the detection problem, the relevant
physics inputs to the model are the
probability distributions of the two orthogonal quadrature components.

According to Weiner's formulation \cite{GC:Weiner88},
in fluctuation class $\kappa$, the scaled target power RV may be characterized as
\begin{equation}
\hat{\mathbf{X}}_{\text{s}}^2 = \frac{1}{\kappa}\sum_{k=1}^\kappa
     \mathbf{H}_k^{\sf T}\mathbf{H}_k \,,
\label{XI}
\end{equation}
where, for each
\mbox{$k = 1,2\ldots,\kappa$},
the vector RV $\mathbf{H}_k$ is a zero-mean unit-variance Gaussian with common correlation
matrix $\mtx{C}_{\text{s}}$.
The associated probability measure
\mbox{$F_\mathbf{H}(d\mathbf{h}) = P_\mathbf{H}(\mathbf{h})d\mathbf{h}$}
is thus given by
\begin{equation}
F_\mathbf{H}(d\mathbf{h}) = \frac{d\mathbf{h}}{\left[(2\pi)^{M}\det\mtx{C}_{\text{s}}\right]^{\kappa/2}}
     \exp\left[-\frac{1}{2}\sum_{k=1}^\kappa
     \mathbf{h}_k^{\sf T}\mtx{C}^{-1}_{\text{s}}\mathbf{h}_k\right] \,,
\label{FI}
\end{equation}
where $d\mathbf{h}$ is the product measure in $\kappa M$ dimensions
\begin{equation}
d\mathbf{h} \equiv \prod_{k = 1}^\kappa \prod_{m = 1}^M dh_{k,m} \;,
\end{equation}
and where $h_{k,m}$ denotes the $m$-th component of the vector $\mathbf{h}_k$.
We observe the explicit orthogonal invariance
\mbox{$F(d(\mtx{R}\mathbf{h}); \mtx{R}\mtx{C}_{\text{s}}\mtx{R}^{\sf T}) = F(d\mathbf{h}; \mtx{C}_{\text{s}})$},
where
\mbox{$F(d\mathbf{h}; \mtx{C}_{\text{s}}) \equiv F_\mathbf{H}(d\mathbf{h})$}
and
\mbox{$\mtx{R}\mathbf{h} \equiv (\mtx{R}\mathbf{h}_1, \mtx{R}\mathbf{h}_2,\ldots,\mtx{R}\mathbf{h}_\kappa)$}.
If we write
\mbox{$T \equiv \hat{\mathbf{X}}_{\text{s}}^2$},
then the induced probability measure
\mbox{$F_T(dt) = P_T(t)dt$}
on the space of functions of the scalar RV $T$ is given by
\begin{equation}
F_T(dt) = \int_{\mathbb{R}^{\kappa M}} d\mathbf{h}\, P_\mathbf{H}(\mathbf{h})
     \delta\Bigl(t - \frac{1}{\kappa}\sum_{k=1}^\kappa \mathbf{h}_k^{\sf T}
     \mathbf{h}_k\Bigr){\cdot}dt \,,
\end{equation}
where
\mbox{$\delta(t)$}
is the Dirac delta function.
Appealing to the Fourier transform of the delta function
\begin{equation}
\delta(t) = \int^{+\infty}_{-\infty}\frac{d\omega}{2\pi}\, e^{i\omega t}  \;,
\end{equation}
it follows immediately that
\begin{equation}
P_{\text{T}}(t) = \int^{+\infty}_{-\infty}\frac{d\omega}{2\pi}\, e^{i\omega t}
     \left[\det\left(\mathbb{I} - \frac{i\omega}{\kappa/2}C_{\text{s}}\right)
     \right]^{-\kappa/2} \,,
\end{equation}
which depends only on the eigenvalues of $\mtx{C}_{\text{s}}$,
as expected due to the invariance under the mapping
\mbox{$\mtx{C}_{\text{s}} \mapsto \mtx{R}\mtx{C}_{\text{s}}\mtx{R}^{\sf T}$}.
When
\mbox{$\mtx{C}_{\text{s}} = \mathbb{I}$},
we recognize, in the integrand, the characteristic function for the distribution
\mbox{$T \sim \Gamma(M\kappa/2,2/\kappa)$}, which corresponds to shape parameter
$M\kappa/2$ and mean $M$.
And if we write
\begin{equation}
T_m \equiv \frac{1}{\kappa}\sum_{k=1}^\kappa H_{mk}^2 \,,
\end{equation}
so that
\mbox{$T = \sum_{m=1}^M T_m$},
then
\mbox{$T_m \sim \Gamma(\kappa/2,2/\kappa)$}
for each
\mbox{$m = 1,2,\ldots,M$}.
The relevant question now is how to extend this discussion in order to construct
a compatible probability measure on the space of functions of the vector RV
$\hat{\mathbf{X}}_{\text{s}}$.

At this point, we can see that
(\ref{KapExp}) may be established by appealing to the orthogonal invariance
of the expectation on the LHS.
Specifically, the expectation
\begin{equation}
\begin{split}
\left\langle\exp\left(-\frac{\sigma^2_{\text{s}}}{2\xi^2}\hat{\mathbf{X}}^2_{\text{s}}\right)
     \right\rangle_{\hat{\mathbf{X}}_{\text{s}}} =
     \left\langle\exp\left(-\frac{\sigma^2_{\text{s}}}{2\kappa\xi^2}
     \sum_{k=1}^\kappa \mathbf{H}_k^{\sf T}\mathbf{H}_k\right)
     \right\rangle_{\mathbf{H}}
\end{split}
\raisetag{-1pt}
\end{equation}
is unchanged under the mapping
\mbox{$\mathbf{H}_k\mapsto \mtx{R}\mathbf{H}_k$}
for any orthogonal matrix $\mtx{R}$, provided that it is accompanied by the mapping
\mbox{$\mtx{C}_{\text{s}} \mapsto \mtx{R}\mtx{C}_{\text{s}}\mtx{R}^{\sf T}$}.
To take advantage of this,
we write the target correlation matrix in diagonalized form as
\mbox{$\mtx{C}_{\text{s}} = \mtx{R}_{\text{s}}^{\sf T}\mtx{\Lambda}_{\text{s}}\mtx{R}_{\text{s}}$},
where
\mbox{$\mtx{\Lambda}_{\text{s}} \equiv \diag(\gamma^{\text{s}}_1,\gamma^{\text{s}}_2,\ldots,\gamma^{\text{s}}_M)$}
and
\mbox{$\mtx{R}_{\text{s}}^{\sf T}\mtx{R}_{\text{s}} = \mathbb{I}$},
and introduce the rotated vector RVs
\mbox{$\mathbf{J}_k \equiv \mtx{R}_{\text{s}}\mathbf{H}_k$}
whose elements are independent zero-mean Gaussians such that the $J_{mk}$ have common
variance $\gamma^{\text{s}}_m$ for
\mbox{$k = 1,2,\ldots,\kappa$}.
It follows that the sums
\begin{equation}
V_m \equiv \frac{1}{\kappa}\sum_{k=1}^\kappa J^2_{mk}
\end{equation}
are independent and gamma-distributed with respective means $\gamma^{\text{s}}_m$.
Then, we explicitly have
\begin{equation}
\left\langle\exp\left(-\frac{\sigma^2_{\text{s}}}{2\xi^2}\sum_{m=1}^M V_m\right)
     \right\rangle_{\mathbf{V}} =
     \prod_{m=1}^M\left(1 + \frac{\sigma^2_{\text{s}}}{\kappa\xi^2}\gamma^{\text{s}}_m
     \right)^{-\kappa/2} \,,
\end{equation}
and this is equivalent to (\ref{KapExp}).

We may apply the same rotational change of variable scheme to the problem with correlated clutter,
where $\mtx{Q}$ is a general matrix, provided we can assume that the, yet unspecified, mapping
\mbox{$(\mathbf{H}_1,\mathbf{H}_2,\ldots,\mathbf{H}_\kappa) \mapsto \hat{\mathbf{X}}_{\text{s}}$}
satisfies\linebreak
\mbox{$(\mtx{R}\mathbf{H}_1,\mtx{R}\mathbf{H}_2,\ldots,\mtx{R}\mathbf{H}_\kappa) \mapsto
     \mtx{R}\hat{\mathbf{X}}_{\text{s}}$}
for any orthogonal matrix $\mtx{R}$.
This requirement is clearly consistent with (\ref{XI}) but its existence for
\mbox{$\kappa > 1$}
is not clear.
In this case, the argument of the expectation in (\ref{XQX})
is not rotationally invariant, but the induced probability measure on $\hat{\mathbf{X}}_{\text{s}}$
will be invariant under the simultaneous mappings
\mbox{$\hat{\mathbf{X}}_{\text{s}} \mapsto \mtx{R}_{\text{s}}\hat{\mathbf{X}}_{\text{s}}$},
\mbox{$C_{\text{s}} \mapsto \mtx{R}_{\text{s}}\mtx{C}_{\text{s}}\mtx{R}_{\text{s}}^{\sf T}$}.
Thus, setting
\mbox{$\mathbf{Y} \equiv \mtx{R}_{\text{s}}\hat{\mathbf{X}}_{\text{s}}$},
we obtain
\begin{align}
\begin{aligned}
&\left\langle\exp\left(-\frac{\sigma^2_{\text{s}}}{2}\hat{\mathbf{X}}_{\text{s}}^{\sf T}\mtx{Q}^{-1}
     \hat{\mathbf{X}}_{\text{s}}\right)
     \right\rangle_{\hat{\mathbf{X}}_{\text{s}}} \\
&\quad = \left\langle\exp\left(-\frac{\sigma^2_{\text{s}}}{2}\hat{\mathbf{X}}_{\text{s}}^{\sf T}\mtx{Q}^{-1}
     \hat{\mathbf{X}}_{\text{s}}\right)
     \right\rangle_{\mtx{R}_{\text{s}}\hat{\mathbf{X}}_{\text{s}}} \\
&\quad =  \left\langle\exp\left(-\frac{\sigma^2_{\text{s}}}{2}\mathbf{Y}^{\sf T}
     (\mtx{R}_{\text{s}}\mtx{Q}^{-1}\mtx{R}_{\text{s}}^{\sf T})\mathbf{Y}\right)
     \right\rangle_{\mathbf{Y}} \\
&\quad =  \left\langle\exp\left(-\frac{\sigma^2_{\text{s}}}{2}\hat{\mathbf{Y}}^{\sf T}
     (\mtx{L}_{\text{s}}\mtx{Q}^{-1}\mtx{L}_{\text{s}}^{\sf T})\hat{\mathbf{Y}}\right)
     \right\rangle_{\hat{\mathbf{Y}}} \,,
\label{LQL}
\end{aligned}
\end{align}
where
\mbox{$\mathbf{Y} \equiv  \sqrt{\mtx{\Lambda}_{\text{s}}}\hat{\mathbf{Y}}$},
and
\mbox{$\mtx{L}_{\text{s}} \equiv \sqrt{\mtx{\Lambda}_{\text{s}}}\mtx{R}_{\text{s}}$}
so that
\mbox{$\mtx{C}_{\text{s}} = \mtx{L}_{\text{s}}^{\sf T}\mtx{L}_{\text{s}}$}.
We note that elements of the vector RV $\hat{\mathbf{Y}}$ are
independent and identically distributed (iid) such that
\mbox{$\hat{Y}^2_m \sim \Gamma(\kappa/2,2/\kappa)$}
for all
\mbox{$m = 1,2\ldots,M$}.

The problem of an uncorrelated gamma-fluctuating target embedded in correlated clutter
is obtained when
\mbox{$\mtx{L}_{\text{s}} = \mathbb{I}$}.
Therefore, we see that introducing target correlation in the presence of correlated
clutter simply amounts to making the mapping
\mbox{$\mtx{Q}^{-1} \mapsto \mtx{L}_{\text{s}}\mtx{Q}^{-1}\mtx{L}_{\text{s}}^{\sf T}$}
in the calculation of the target expectation.
When
\mbox{$\mtx{C}_{\text{s}} = \mathbb{I}$},
the only sensible option is for
each of iid RVs $\hat{X}_m$ to be described by a two-sided Nakagami distribution
whose PDF is given by
\begin{equation}
\begin{split}
&f(x; \nu,\Omega)dx = {} \\
&    \frac{1}{\Gamma(\nu)}e^{-\left(x\sqrt{\nu/\Omega}\right)^2}
     \left(|x|\sqrt{\nu/\Omega}\right)^{2\nu-1}d(x\sqrt{\nu/\Omega}) \,,
\label{Nak}
\end{split}
\end{equation}
with fading parameter
\mbox{$\nu = \kappa/2$}
and scale parameter
\mbox{$\Omega = 1$},
and this becomes Gaussian for
\mbox{$\kappa = 1$}.
The characteristic function is given by \cite{GC:Simon05}
\begin{align}
\begin{aligned}
\tilde{f}(u;\nu,\Omega) &\equiv \left\langle e^{-iu\hat{X}}\right\rangle_{\hat{X}}
     = \frac{\Gamma(2\nu)}{2^\nu\Gamma(\nu)}\exp\left(-\frac{\Omega}{8\nu}u^2\right) \\
& {}\times \left[D_{-2\nu}\left(iu\sqrt{\frac{\Omega}{2\nu}}\right) +
     D_{-2\nu}\left(-iu\sqrt{\frac{\Omega}{2\nu}}\right)\right] \,,
\end{aligned}
\end{align}
where
\mbox{$D_{-\nu}(z)$}
is a parabolic cylinder function \cite{GC:Gradshteyn07}.
Regardless of whether it is possible to find a suitable rotationally
covariant mapping
\mbox{$\{\mathbf{H}_\kappa\} \mapsto \mathbf{X}_{\text{s}}$},
we shall adopt (\ref{LQL}) as the {\em definition} of a
candidate first-principles model: that is, we set
\mbox{$\hat{\mathbf{X}}_{\text{s}} \equiv \mtx{L}^{\sf T}_{\text{s}}\hat{\mathbf{Y}}$},
and furthermore postulate that, even
for a general target correlation matrix $\mtx{C}_{\text{s}}$, the components of
the vector RV $\hat{\mathbf{Y}}$ are iid and generated by the two-sided
Nakagami distribution.

\subsection{Statistical Properties of the First-Principles Model}
The aim of this section is to demonstrate that the model proposed in the
foregoing section yields the required mean and variance as given
by (\ref{MnVar1})--(\ref{MnVar3}).
Recalling that the mean and variance are respectively given
by the first and second derivatives of the cumulant generating function (CGF)
evaluated at the origin, we proceed by
writing down the CGF $\CGF(s)$ for the model and making a Taylor expansion about
\mbox{$s = 0$}
up to second order.

Thus, in order to compute the mean and variance arising from (\ref{XQX}),
let us introduce the functions
\begin{align}
\begin{aligned}
\CGF_0(s) &\equiv -\ln\det(s{\cdot}\mtx{Q}(s)) \,, \\
\CGF_1(s) &\equiv 2\ln\left\langle\exp\left(
     -\frac{\sigma^2_{\text{s}}}{2}\hat{\mathbf{X}}_{\text{s}}^{\sf T}\mtx{Q}^{-1}(s)
     \hat{\mathbf{X}}_{\text{s}}\right)
     \right\rangle_{\hat{\mathbf{X}}_{\text{s}}} \,.
\end{aligned}
\end{align}
Then, the cumulant generating function for the model defined by (\ref{XQX})
is given by
\begin{equation}
\CGF(s) \;\equiv\; \ln\left\langle e^{-sZ_{\text{avg}}} \right\rangle_{Z_{\text{avg}}}
     \; = \; \CGF_0(s) + \CGF_1(s) \,.
\end{equation}
The function $\CGF_0(s)$ generates the contribution to the CGF
that is independent of the SIR. Thus,
performing a Taylor expansion in $s$,
\begin{equation}
\begin{split}
\CGF_0(s) = &-s(\sigma^2_{\text{n}} + \sigma^2_{\text{c}}) + \frac{s^2}{2}
     \biggl[\frac{\sigma^4_{\text{n}}}{M} + \frac{\sigma^4_{\text{c}}}{M^2}\Tr\{\mtx{C}_{\text{c}}^2\}
     + \frac{2\sigma^2_{\text{n}}\sigma^2_{\text{c}}}{M}\biggr] \\
&    {}+ O(s^3).
\end{split}
\end{equation}
Also performing a Taylor expansion in $s$ for the SIR-dependent term yields
\begin{equation}
\begin{split}
&\CGF_1(s) = -\sigma^2_{\text{s}}\left\langle\hat{\mathbf{X}}_{\text{s}}^{\sf T}\mtx{Q}^{-1}(s)
     \hat{\mathbf{X}}_{\text{s}}\right\rangle_{\hat{\mathbf{X}}_{\text{s}}} \\
& {}+ \frac{\sigma^4_{\text{s}}}{4}\left[\left\langle\left(\hat{\mathbf{X}}_{\text{s}}^{\sf T}\mtx{Q}^{-1}(s)
     \hat{\mathbf{X}}_{\text{s}}\right)^2\right\rangle_{\hat{\mathbf{X}}_{\text{s}}} -
     \left(\left\langle\hat{\mathbf{X}}_{\text{s}}^{\sf T}\mtx{Q}^{-1}(s)
     \hat{\mathbf{X}}_{\text{s}}\right\rangle_{\hat{\mathbf{X}}_{\text{s}}}\right)^2\right] \\
& {}+ O(s^3) \,,
\end{split}
\end{equation}
and we note that
\begin{align}
\begin{aligned}
\mtx{Q}^{-1}(s) &= \frac{s}{1 + s(\sigma^2_{\text{n}}\mathbb{I} + \sigma^2_{\text{c}}\mtx{C}_{\text{c}})} \\
&= s\mathbb{I} - s^2(\sigma^2_{\text{n}}\mathbb{I} + \sigma^2_{\text{c}}\mtx{C}_{\text{c}})  + O(s^3) \,,
\end{aligned}
\label{Qinv}
\end{align}
which yields
\begin{equation}
\begin{split}
&\CGF_1(s) = -s\sigma^2_{\text{s}}\left\langle\hat{\mathbf{X}}_{\text{s}}^{\sf T}
     \hat{\mathbf{X}}_{\text{s}}\right\rangle_{\hat{\mathbf{X}}_{\text{s}}}
     + s^2\biggl[\sigma^2_{\text{n}}\sigma^2_{\text{s}}
     \left\langle\hat{\mathbf{X}}_{\text{s}}^{\sf T}
     \hat{\mathbf{X}}_{\text{s}}\right\rangle_{\hat{\mathbf{X}}_{\text{s}}} \\
& {}+ \sigma^2_{\text{c}}\sigma^2_{\text{s}}
     \left\langle\hat{\mathbf{X}}_{\text{s}}^{\sf T}\mtx{C}_{\text{c}}\hat{\mathbf{X}}_{\text{s}}
     \right\rangle_{\hat{\mathbf{X}}_{\text{s}}} + \frac{\sigma^2_{\text{s}}}{4}
     \var(\hat{\mathbf{X}}_{\text{s}}^{\sf T}\hat{\mathbf{X}}_{\text{s}})\biggr] + O(s^3) \,.
\end{split}
\end{equation}
A matrix in the denominator, as in (\ref{Qinv}), denotes the inverse.

By construction, (\ref{XI}) holds in distribution for the first-principles model.
This is easily confirmed, based on the observation that
\begin{equation}
H \sim \mathcal{N}(0,1) \Rightarrow H^2 \sim \Gamma(1/2, 2) \,.
\end{equation}
It follows immediately that
\begin{equation}
\left\langle\hat{\mathbf{X}}_{\text{s}}^{\sf T}
     \hat{\mathbf{X}}_{\text{s}}\right\rangle_{\hat{\mathbf{X}}_{\text{s}}}
     = \Tr\{\mtx{C}_{\text{s}}\} \,, \quad
\var(\hat{\mathbf{X}}_{\text{s}}^{\sf T}\hat{\mathbf{X}}_{\text{s}})
     = \frac{2}{\kappa}\Tr\{\mtx{C}_{\text{s}}^2\} \,.
\end{equation}
From this, we see that the mean and variance are given as in
(\ref{MnVar1})--(\ref{MnVar3}), but with
\begin{equation}
N = M^2/\left\langle\hat{\mathbf{X}}_{\text{s}}^{\sf T}\mtx{C}_{\text{c}}\hat{\mathbf{X}}_{\text{s}}
     \right\rangle_{\hat{\mathbf{X}}_{\text{s}}} \,.
\end{equation}
To compute this expectation, we recall
that the RVs $\hat{Y}_m$,
\mbox{$m = 1,2,\ldots,M$}
are iid such that
\mbox{$\hat{Y}_m^2 \sim \Gamma(\kappa/2,2/\kappa)$}
and with random signs ({\it i.e.}\ randomly positive or negative, so that
\mbox{$\sgn(\hat{Y}_m) = \pm 1$}
is by construction a Rademacher RV), and set
\mbox{$\hat{\mathbf{X}}_{\text{s}} \equiv \mtx{L}^{\sf T}_{\text{s}}\hat{\mathbf{Y}}$},
where
\mbox{$\mtx{C}_{\text{s}} = \mtx{L}^{\sf T}_{\text{s}}\mtx{L}_{\text{s}}$}.
Accordingly,
\begin{align}
\begin{aligned}
\left\langle\hat{X}_m\hat{X}_n\right\rangle_{\hat{\mathbf{X}}_{\text{s}}}
&= \sum_{a,b=1}^M \left[\mtx{L}_{\text{s}}\right]_{am}\left[\mtx{L}_{\text{s}}\right]_{bn}
     \left\langle \hat{Y}_a\hat{Y}_b\right\rangle_{\mathbf{Y}} \\
& = \sum_{a,b=1}^M \left[\mtx{L}_{\text{s}}\right]_{am}\left[\mtx{L}_{\text{s}}\right]_{bn}
     \delta_{ab} \\
&= \left[\mtx{L}^{\sf T}_{\text{s}}\mtx{L}_{\text{s}}\right]_{mn} \\
&= \left[\mtx{C}_{\text{s}}\right]_{mn} \,.
\end{aligned}
\end{align}
This means that we obtain
\mbox{$\left\langle\hat{\mathbf{X}}_{\text{s}}^{\sf T}\mtx{C}_{\text{c}}\hat{\mathbf{X}}_{\text{s}}
     \right\rangle_{\hat{\mathbf{X}}_{\text{s}}} = \Tr\{\mtx{C}_{\text{c}}\mtx{C}_{\text{s}}\}$}
as desired.
The upshot of all this is that the candidate first-principles model reproduces
the expected mean and variance for the returned power, and (\ref{LQL}) holds,
but the rotational covariance assumption that initially led to it is redundant.

\subsection{Model Summary}
In summary, the first-principles model is defined by specifying that the
(normalized) target quadrature component is given by
\mbox{$\hat{\mathbf{X}}_{\text{s}} \equiv \mtx{L}^{\sf T}_{\text{s}}\hat{\mathbf{Y}}$},
where the iid vector RV $\hat{\mathbf{Y}}$ has elements distributed according
to the two-sided Nakagami distribution of (\ref{Nak}),
and the matrix $\mtx{L}_{\text{s}}$ is obtained from the eigenvalue decomposition of
the target correlation matrix according to
\begin{equation}
\mtx{C}_{\text{s}} = \mtx{R}_{\text{s}}^{\sf T}\mtx{\Lambda}_{\text{s}}\mtx{R}_{\text{s}} \,, \quad
\mtx{L}_{\text{s}} \equiv \sqrt{\mtx{\Lambda}_{\text{s}}}\mtx{R}_{\text{s}} \;.
\end{equation}
Hence, the MGF of the first-principles model is given by
\begin{equation}
\begin{split}
&\MGF_Z(s) = {} \\
&    \frac{1}{s^M\det\mtx{Q}(s)}{\cdot}
     \left\langle\exp\left(-\frac{\sigma^2_{\text{s}}}{2}\hat{\mathbf{Y}}^{\sf T}
     (\mtx{L}_{\text{s}}\mtx{Q}^{-1}(s)\mtx{L}_{\text{s}}^{\sf T})\hat{\mathbf{Y}}\right)
     \right\rangle^2_{\hat{\mathbf{Y}}} \,.
\label{LZFP}
\end{split}
\end{equation}

In the Appendix, we discuss some minor shortcomings of the first-principles model.
We also show that it may be computed exactly for the special case of a fully
correlated target.

Compound clutter may be incorporated as described at the end of Section~\ref{ProbForm}
by making the Q-matrix of (\ref{Qmatrix}) dependent on the clutter texture integration
variable $u$, according to
\begin{equation}
\mtx{Q}(s; u) \equiv (1/s + \sigma^2_{\text{n}})\mathbb{I} +
     u\bar{\sigma}^2_{\text{c}}\mtx{C}_{\text{c}} \,,
\end{equation}
or, equivalently, the normalized form
\begin{equation}
\mtx{Q}(s; u)/\sigma_{\text{I}}^2 = \left(\frac{1}{s\sigma_{\text{I}}^2} +
     1 - q\right)\mathbb{I} + qu\mtx{C}_{\text{c}} \,.
\end{equation}
The calculations proceed for an arbitrary value of $u$, until the texture integration
is performed at the end, as indicated in (\ref{Fspk}).

\section{Effective Model}
\label{EffModel}
A `first-principles' model for the problem of a correlated gamma-fluctuating target
in correlated clutter would be characterized by the specification of the joint
probability distribution for the quadrature RVs $\hat{\mathbf{X}}_{\text{s}}$
that appear in (\ref{XQX}) such that (\ref{LQL}) holds.
However, having such a joint probability distribution
does not necessarily, or even likely, lead to a convenient closed-form
MGF to which saddle-point techniques could be easily applied.
This is the case with the proposal of the preceding section,
as defined by (\ref{LZFP}), which appears to be analytically intractable.
So in this section, we propose an alternative `effective' model --- one whose
outputs are expected to be that same as those of the first-principles model,
though not specified in terms of physical degrees of freedom ---
that is amenable to analytical treatment.
Specifically, our approach is to postulate an explicit MGF that reduces to the
expected results in all known computable special cases. This MGF would then define
an `effective model' for the system.
We examine
its relationship with the previously defined first-principles model,
show how to compute the survival function from the MGF using
saddle-point techniques and, finally,
present two approximation schemes that reduce computational load
significantly while maintaining a high degree of accuracy.

Let us consider the following `effective' MGF for the problem with correlated clutter:
\begin{equation}
\MGF_{Z_{\text{avg}}}(s) = \frac{\left[\det\left(\mathbb{I} + (s/M)\left(\sigma^2_{\text{n}}
     + \sigma^2_{\text{c}}\mtx{C}_{\text{c}}\right)\right)\right]^{\kappa-1}}
     {\left[\det\left(\mathbb{I} + (s/M)\left(\sigma^2_{\text{n}}
     + \sigma^2_{\text{c}}\mtx{C}_{\text{c}} + \sigma^2_{\text{s}}\mtx{C}_{\text{s}}/\kappa
     \right)\right)\right]^\kappa} \,,
\label{Leff}
\end{equation}
which correctly reproduces the special cases (i)
\mbox{$\sigma^2_{\text{c}} = 0$},
(ii)
\mbox{$\sigma^2_{\text{s}} = 0$},
(iii)
\mbox{$\rho_{\text{c}} = 0$},
and (iv)
\mbox{$\kappa = 1$}.
It also satisfies the essential requirement of yielding the expected mean
and variance, as given by (\ref{MnVar1})--(\ref{MnVar3}).
To see that, we construct the CGF
\begin{equation}
\CGF(t) \;\equiv\; \ln \MGF_{Z_{\text{avg}}}(t) \; = \;
     \ln\left\langle e^{-tZ_{\text{avg}}} \right\rangle_{Z_{\text{avg}}} \,,
\end{equation}
which, for the present problem, reads
\begin{align}
\begin{aligned}
\CGF(t) &= (\kappa-1)\Tr\left\{\ln\left[\mathbb{I} + (t/M)(\sigma^2_{\text{n}} +
     \sigma^2_{\text{c}}\mtx{C}_{\text{c}})\right]\right\} \\
&\quad {}- \kappa\Tr\left\{\ln\left[\mathbb{I} +
     (t/M)(\sigma^2_{\text{n}} + \sigma^2_{\text{c}}\mtx{C}_{\text{c}} +
     \sigma^2_{\text{s}}\mtx{C}_{\text{s}}/\kappa)\right]\right\} \\
& = -t(\sigma^2_{\text{n}} + \sigma^2_{\text{c}} + \sigma^2_{\text{s}}) \\
&\quad {}+ \frac{t^2}{2}
     \biggl[\frac{\sigma^4_{\text{n}}}{M} + \frac{\sigma^4_{\text{c}}}{M^2}\Tr\{\mtx{C}_{\text{c}}^2\}
     + \frac{\sigma^4_{\text{s}}}{\kappa M^2}\Tr\{\mtx{C}_{\text{s}}^2\} \\
&\quad {}+2\left(\frac{\sigma^2_{\text{n}}\sigma^2_{\text{c}}}{M} + \frac{\sigma^2_{\text{n}}\sigma^2_{\text{s}}}{M}
    + \frac{\sigma^2_{\text{c}}\sigma^2_{\text{s}}}{M}\Tr\{\mtx{C}_{\text{c}}\mtx{C}_{\text{s}}\}\right)\biggr]
    + O(t^3) \,,
\end{aligned}
\end{align}
and recall that the mean and variance are given by the first and second derivatives of the
CGF evaluated at the origin.
This expression leads directly to the results given in (\ref{MnVar1}) and (\ref{MnVar2})
for the returned power mean and variance, respectively, noting that we must set
\mbox{$\nu = \infty$}
for the clutter shape parameter
as we have not yet introduced the compound clutter texture.
The form of the MGF in (\ref{Leff}) is an ansatz ({\it i.e.}\ trial form)
motivated by previous work on the
heuristic generalized Dalle Mese Giuli (DMG) approximation \cite{GC:Zuk18}.

In order to make contact with the previous approach, we can write
\begin{align}
\begin{aligned}
\MGF_Z(s) &= \frac{1}{s^M\det\mtx{Q}(s)}\left[\frac{\det\mtx{Q}(s)}{\det\left(\mtx{Q}(s) +
     \sigma^2_{\text{s}}\mtx{C}_{\text{s}}/\kappa\right)}\right]^\kappa \\
&= \frac{1}{s^M\det\mtx{Q}(s)}\left[\det\left(\mathbb{I} + \frac{\sigma^2_{\text{s}}}{\kappa}
     \mtx{L}_{\text{s}}\mtx{Q}^{-1}(s)\mtx{L}_{\text{s}}^{\sf T}\right)\right]^{-\kappa} \,.
\end{aligned}
\label{LZEff}
\end{align}
We see that
reconciling the first-principles approach of (\ref{XQX}) and (\ref{LQL})
with the effective model of (\ref{Leff}) would require that
\begin{equation}
\left\langle\exp\left(-\tfrac{1}{2}\hat{\mathbf{Y}}^{\sf T}\mtx{A}\hat{\mathbf{Y}}\right)
     \right\rangle_{\hat{\mathbf{Y}}} =
     \left[\det\left(\mathbb{I} + \mtx{A}/\kappa\right)\right]^{-\kappa/2} \,,
\end{equation}
for any positive semi-definite symmetric matrix $\mtx{A}$.
We should note that this statement does indeed hold for any such diagonal matrix $\mtx{A}$,
or when
\mbox{$\kappa = 1$}
in which case $\hat{\mathbf{Y}}$
is a Gaussian RV.
However, this relationship will not hold in general: The RHS manifestly only depends
on the eigenvalues of $\mtx{A}$. The LHS will depend only on the eigenvalues if the probability
measure on $\hat{\mathbf{Y}}$ is rotationally invariant, which is not expected to
be that case for
\mbox{$\kappa > 1$}.
Nevertheless, it is useful to consider the special case where,
as is true in the generalized DMG model \cite{GC:Zuk18}
(discussed in more detail in a later section),
the matrices $\mtx{C}_{\text{c}}$ and $\mtx{C}_{\text{s}}$ commute, and are thus simultaneously
diagonalizable. Then, upon writing
\mbox{$\mtx{C}_{\text{c}} = \mtx{R}^{-1}\mtx{\Lambda}_{\text{c}}\mtx{R}$}
and
\mbox{$\mtx{C}_{\text{s}} = \mtx{R}^{-1}\mtx{\Lambda}_{\text{s}}\mtx{R}$}
so that
\mbox{$\mtx{L}_{\text{s}} = \sqrt{\mtx{\Lambda}_{\text{s}}}\mtx{R}$},
we have
\begin{equation}
\mtx{A} \; = \; \sigma^2_{\text{s}}\mtx{L}_{\text{s}}\mtx{Q}^{-1}(s)\mtx{L}_{\text{s}}^{\sf T}
\; = \; \frac{\sigma^2_{\text{s}}\mtx{\Lambda}_{\text{s}}}{1/s + \sigma^2_{\text{n}}\mathbb{I}
     + \sigma^2_{\text{c}}\mtx{\Lambda}_{\text{c}}} \,,
\label{commute}
\end{equation}
which is diagonal, and so the equivalence holds.
It follows that the DMG model
is simultaneously an approximation to the effective model
and any valid first-principles model.

It is also true, in the general case,
that the correlation matrices $\mtx{C}_{\text{c}}$ and $\mtx{C}_{\text{s}}$
commute asymptotically in the limit as
\mbox{$M\to\infty$},
in the sense that the weak matrix norm of the commutator vanishes in this limit,
provided that they are symmetric Toeplitz and satisfy the summability
conditions \cite{GC:Gray06}
\begin{equation}
\sum_{m = 1}^\infty \left|[\mtx{C}_{\text{c}}]_{m1}\right| < \infty \,, \quad
\sum_{m = 1}^\infty \left|[\mtx{C}_{\text{s}}]_{m1}\right| < \infty \,.
\end{equation}
Thus, (\ref{commute}) holds asymptotically, from which it follows that the
effective model is (at least) asymptotically equivalent to the first-principles model.
Finally, we remark that, when
\mbox{$M = 2$},
all correlation matrices will have the circulant form
\begin{equation}
\mtx{C} =
\begin{pmatrix}
1    & \rho \\
\rho & 1
\end{pmatrix} \,,
\end{equation}
and therefore all such matrices commute.
We conclude that the first-principles and effective models must
be identical for
\mbox{$M = 2$}.

\subsection{Saddle-point Method for the Effective Model}
A robust and accurate way of computing the survival function for a distribution
from its MGF is provided by saddle-point techniques, which are especially
effective when the MGF is a rational function.
This approach has previously been developed for the case of a correlated
gamma-fluctuating target embedded in uncorrelated clutter speckle \cite{GC:Zuk19},
and is easily extended to the general problem.

In the presence of clutter texture, the speckle MGF for the effective model can be
expressed as
\begin{equation}
\begin{split}
\MGF^{\text{spk}}_{Z'_{\text{avg}}}(s;u) = \frac{\left[\det\left(\mathbb{I}
     + (s/M)(1-q + qu\mtx{C}_{\text{c}})\right)\right]^{\kappa-1}}
     {\left[\det\left(\mathbb{I}
     + (s/M)(1-q + qu\mtx{C}_{\text{c}} + S\mtx{C}_{\text{s}}/\kappa)\right)\right]^{\kappa}} \,.
\end{split}
\raisetag{-1pt}
\end{equation}
where $u$ denotes the normalized texture integration variable.
It is convenient to introduce an aggregated target/clutter correlation matrix according to
\begin{equation}
\mtx{C}_{\text{sc}}(u) \equiv \frac{1}{qu + S/\kappa}\left[qu\mtx{C}_{\text{c}} +
     (S/\kappa)\mtx{C}_{\text{s}}\right] \,,
\end{equation}
such that
\mbox{$\Tr\mtx{C}_{\text{sc}} = M$}.
Then, we can write
\begin{equation}
\MGF^{\text{spk}}_{Z'_{\text{avg}}}(s;u) = \frac{\prod_{m=1}^M\left[1 + a_{qm}(u)s\right]^{\kappa-1}}
     {\prod_{m=1}^M\left[1 + a_m(u)s\right]^\kappa} \,,
\label{mgfaq}
\end{equation}
with
\begin{align}
\begin{aligned}
a_{qm}(u) &= \left[1 - q + qu\gamma^{\text{c}}_m\right]/M \,, \\
a_m(u)    &= \left[1 - q + (qu + S/\kappa)\gamma^{\text{sc}}_m(u)\right]/M \,,
\end{aligned}
\label{AFromGam}
\end{align}
where
\begin{align}
\begin{aligned}
\{\gamma^{\text{c}}_m: m = 1,2,\ldots,M\} &= \eig(\mtx{C}_{\text{c}}) \,, \\
\{\gamma^{\text{sc}}_m(u): m = 1,2,\ldots,M\} &= \eig(\mtx{C}_{\text{sc}}(u)) \,.
\end{aligned}
\label{GamEig}
\end{align}
It is useful to note that
\begin{equation}
\sum_{m=1}^M (a_m - a_{qm}) = S/\kappa \,.
\end{equation}

The saddle-point equation, for the saddle-point (SP)
\mbox{$s = s_0$},
is obtained as the extremum
(\mbox{$\Phi'(s) = 0$})
of Helstrom's phase function \cite{GC:Helstrom92}
\begin{equation}
\Phi(s;v,u) \equiv \ln \MGF^{\text{spk}}_{Z'_{\text{avg}}}(s;u) - \ln(-s) + sv \,,
\label{PhiPhase}
\end{equation}
and reads
\begin{equation}
v = \frac{1}{s_0} + \kappa\sum_{m=1}^M \frac{a_m(u)}{1 + a_m(u)s_0}
     - (\kappa-1)\frac{a_{qm}(u)}{1 + a_{qm}(u)s_0} \,.
\end{equation}
The integration contour for the inverse Laplace transform
of the MGF that yields
the survival function is taken
to be the steepest descent path (SDP), which is defined by
\mbox{$\Imz\Phi(s) = 0$}
and passes through the saddle-point $s_0$.
Next, for convenience, we set
\mbox{$a_0 \equiv a_{q0} \equiv +\infty$}
and define
\begin{equation}
c_m \equiv \frac{1}{s_0v + v/a_m(u)} \,, \quad
     c_{qm} \equiv \frac{1}{s_0v + v/a_{qm}(u)} \,,
\end{equation}
for
\mbox{$m = 0,1,\ldots,M$}.
Then, with
\mbox{$z \equiv (s_0 - s)v$},
we have the associated $\tau$-phase function \cite{GC:Zuk19}
defined by
\begin{equation}
\tau(z) \equiv \Phi(s_0; v) - \Phi(s_0 - z/v; v) \,,
\end{equation}
and evaluated as
\begin{align}
\begin{aligned}
\tau(z) &= z + \kappa\sum_{m=0}^M\ln(1 - c_mz) -
     (\kappa-1)\sum_{m=0}^M\ln(1 - c_{qm}z) \\
&= -\sum_{n=2}^\infty \frac{r_n}{n} z^n \,,
\label{TauPhase}
\end{aligned}
\end{align}
with coefficients
\begin{equation}
r_n \equiv \sum_{m=0}^M\left[\kappa c^n_m - (\kappa-1)c^n_{qm}\right] \,,
\label{rcoefs}
\end{equation}
for
\mbox{$n = 1,2,\ldots$},
arising in its Taylor expansion.
Inversion of this equation (to obtain
\mbox{$z(\tau)$}),
followed by integration over $\tau$ leads to
the exact representation of the speckle survival function
\begin{equation}
\bar{F}_{\text{spk}}(v; qu) = \bar{F}_{\text{sp}}(v; qu){\cdot}
     \sqrt{\frac{2r_2}{\pi}}\int_0^\infty d\tau\, e^{-\tau}\Imz z(\tau) \,,
\label{Ftau}
\end{equation}
assuming that
\mbox{$s_0 < 0$}\footnote{
The adjustments necessary for positive saddle-points
\mbox{$s_0 > 0$}
and the choice of saddle-point sign are described in
\cite{GC:Helstrom92,GC:Zuk19}.},
with the factor
\begin{equation}
\bar{F}_{\text{sp}}(v; qu) = \frac{1}{v\sqrt{2\pi r_2}}{\cdot}e^{\Phi(s_0; v,u)}
\label{Fsp}
\end{equation}
constituting the basic saddle-point approximation.
It only remains to perform the clutter texture integration to obtain the full
survival function
\begin{equation}
\bar{F}_{\text{cc}}(v) \simeq \sum_{\ell=1}^L w_\ell \bar{F}_{\text{spk}}(v;qu_\ell) \,,
\end{equation}
where the $w_\ell$, $u_\ell$ are the weights and nodes, respectively, of an
appropriate Gaussian quadrature of order $L$.

The Pad\'{e}-adapted version of the $\tau$-phase function reads
\begin{equation}
\begin{split}
\tau(z) &= z + \ln(1 - c_0z) \\
&     {}+ \kappa\left[(M-1)\ln(1 - \bar{c}z) + \ln(1 - c_Mz)\right] \\
&     {}- (\kappa -1)\left[(M-1)\ln(1 - \bar{c}_qz) + \ln(1 - c_{qM}z)\right] \\
&     {}- \sum_{n=2}^\infty\frac{\bar{r}_n}{n} z^n \,,
\end{split}
\label{TauPade}
\end{equation}
with
\begin{equation}
\bar{c} \equiv \frac{1}{M-1}\sum_{m=1}^{M-1} c_m \,, \quad
     \bar{c}_q \equiv \frac{1}{M-1}\sum_{m=1}^{M-1} c_{qm} \,,
\end{equation}
and
\begin{equation}
\bar{r}_n \equiv \sum_{m=1}^{M-1}\left[\kappa(c_m^n - \bar{c}^n) - (\kappa-1)
     (c_{qm}^n - \bar{c}_q^n)\right] \,.
\end{equation}
In the special case of uncorrelated clutter speckle ({\it i.e.}\ the problem
previously studied in \cite{GC:Zuk19}), we have
\mbox{$c_{qm} = \bar{c}_q$}
for all
\mbox{$m = 1,2,\ldots,M$}.
In this case, (\ref{TauPade}) reduces to equation~(50) of \cite{GC:Zuk19}.
Also, the SP equation implies that
\mbox{$r_1 = 1$},
which expands to the relationship
\begin{equation}
c_0 + \kappa\left[c_m + (M-1)\bar{c}\right] - (\kappa-1)\left[
     c_{qM} + (M-1)\bar{c}_q\right] = 1 \,.
\end{equation}
Using this, we can confirm that the contribution to the RHS of (\ref{TauPade})
linear in $z$ vanishes, as expected.

The Pad\'{e} approximation comprises approximating the infinite series in
(\ref{TauPade}) by a low order
Pad\'{e} approximant, as described in \cite{GC:Zuk19}.
This serves to greatly reduce
the computational effort required to perform the inversion $z(\tau)$ implicit in
the integral of (\ref{Ftau}) for very little loss in accuracy. It obviates the need
for the large summations that appear in the first equality of (\ref{TauPhase})
to be performed at every iteration of the complex-plane Newton-Raphson root finding.

The integration in (\ref{Ftau}) is equivalent to integration over the steepest
descent path (SDP) of the phase function (\ref{PhiPhase})
and leads to a numerically exact result.
On the other hand, depending on the accuracy required,
the coefficients $r_n$ of (\ref{rcoefs}) may be used to
construct the first few terms of the residuum series, of which the basic
saddle-point (SP) approximation, given by (\ref{Fsp}), is the leading contribution.

One should note that, when this technique is used in a compound clutter problem,
because the eigenvalues $\gamma^{\text{sc}}_m$
in (\ref{AFromGam}) depend on the texture integration variable $u$,
a separate eigenvalue problem must be solved at each step of the numerical integration.
We proceed to describe some approximation schemes that avoid this complication.

\subsection{Dalle Mese Giuli (DMG) Approximation}
The generalized DMG approximation, as described in \cite{GC:Zuk18},
is based on the use of simplified correlation matrices, chosen to reproduce
the effective number of looks corresponding to the given
correlation coefficients \cite{GC:Rosen17}.
It has already been found to work well in the case of correlated gamma-fluctuating
targets embedded in uncorrelated clutter speckle \cite{GC:Zuk18}.
In that application, its adoption reduced computation time relative to the
steepest-descent-path integration of the full effective model by a significant factor
(primarily as all the $M$-fold summations over pulses are eliminated).
It also provided the basis for proposing the MGF of the effective model.
Furthermore, as discussed above, in addition to being an approximation to the effective
model, it is also a {\it bone fide}\ approximation to the first-principles model.

In the generalized DMG approximation,
the target and clutter correlation matrices
$\mtx{C}_{\text{s}}$, $\mtx{C}_{\text{c}}$
are constructed such that they commute, and are therefore simultaneously diagonalizable.
Consequently, the speckle MGF of (\ref{mgfaq}) holds with
\begin{align}
\begin{aligned}
a_{qm}(u) &= \left(1 - q + qu\gamma^{\text{c}}_m\right)/M \,, \\
a_m(u)    &= a_{qm}(u) + S\gamma^{\text{s}}_m/(\kappa M) \,,
\end{aligned}
\end{align}
where
\begin{align}
\begin{aligned}
\{\gamma^{\text{c}}_m: m = 1,2,\ldots,M\} &= \eig(\mtx{C}_{\text{c}}) \,, \\
\{\gamma^{\text{s}}_m: m = 1,2,\ldots,M\} &= \eig(\mtx{C}_{\text{s}}) \,.
\end{aligned}
\end{align}
Specifically,
\begin{align}
\begin{aligned}
\gamma^{\text{c}}_m = 1 - \rho'_{\text{c}} + M\rho'_{\text{c}}\delta_{mM} \,, \\
\gamma^{\text{s}}_m = 1 - \rho'_{\text{s}} + M\rho'_{\text{s}}\delta_{mM} \,,
\label{gamsc}
\end{aligned}
\end{align}
where $\delta_{mn}$ denotes the Kronecker delta.
The eigenvalues are expressed in terms of effective correlation coefficients, given by
\begin{align}
\begin{aligned}
\rho'_{\text{s}} &= \sqrt{(M/B-1)/(M-1)} \,, \nonumber \\
\rho'_{\text{c}} &= \sqrt{(M/L-1)/(M-1)} \,,
\end{aligned}
\end{align}
where the quantities $B$, $L$ represent the effective numbers of looks
on the target and clutter, respectively. These, in turn, can be obtained
from the true correlation coefficients associated with an assumed correlation
structure. For Kanter's Gauss-Markov correlation model, one has
\begin{align}
\begin{aligned}
B &= M{\cdot}\left[1 + \frac{2\rho^2_{\text{s}}}{1-\rho^2_{\text{s}}}\left(
     1 - \frac{1}{M}{\cdot}\frac{1-\rho^{2M}_{\text{s}}}{1-\rho^2_{\text{s}}}
     \right)\right]^{-1} \,, \\
L &= M{\cdot}\left[1 + \frac{2\rho^2_{\text{c}}}{1-\rho^2_{\text{c}}}\left(
     1 - \frac{1}{M}{\cdot}\frac{1-\rho^{2M}_{\text{c}}}{1-\rho^2_{\text{c}}}
     \right)\right]^{-1} \,.
\end{aligned}
\end{align}
All of this ensures that the DMG approximation reproduces the
power variance of the underlying model to the extent possible
within an effective-looks approach.

From (\ref{gamsc}), we see that
\mbox{$c_m = \bar{c}$}
and
\mbox{$c_{qm} = \bar{c}_q$}
for all
\mbox{$m = 1,2,\ldots,M-1$},
so that the $\tau$-phase function reads
\begin{equation}
\begin{split}
\tau(z) &= z + \ln(1 - c_0z) \\
&     {}+ \kappa\left[(M-1)\ln(1 - \bar{c}z) + \ln(1 - c_Mz)\right] \\
&     {}- (\kappa -1)\left[(M-1)\ln(1 - \bar{c}_qz) + \ln(1 - c_{qM}z)\right] \,.
\end{split}
\label{TauDMG}
\end{equation}
No Pad\'{e} approximation is necessary in this case, as
\mbox{$\bar{r}_n = 0$}
for all
\mbox{$n = 1,2,\ldots$}.
The SP equation reads
\begin{equation}
\begin{split}
v& = \frac{1}{s_0} + \kappa\left[\frac{(M-1)a_1(u)}{1 + a_1(u)s_0} +
     \frac{a_M(u)}{1 + a_M(u)s_0}\right] \\
&    {}- (\kappa-1)\left[
     \frac{(M-1)a_{q1}(u)}{1 + a_{q1}(u)s_0} +
     \frac{a_{qM}(u)}{1 + a_{qM}(u)s_0}\right] \,.
\end{split}
\end{equation}
Its solution is equivalent to finding the roots of a quintic polynomial.
We may also note that, here,
\begin{equation}
r_n = c_0^n + \kappa\left[(M-1)\bar{c}^n + c_M^n\right] - (\kappa-1)
     \left[(M-1)\bar{c}_q^n + c_{qM}^n\right] \,,
\end{equation}
for
\mbox{$n = 1,2,\ldots$},
with the identity
\mbox{$r_1 \equiv 1$}
following from the SP equation.

\begin{figure}[!t]
\centering
\includegraphics[width=3.45in]{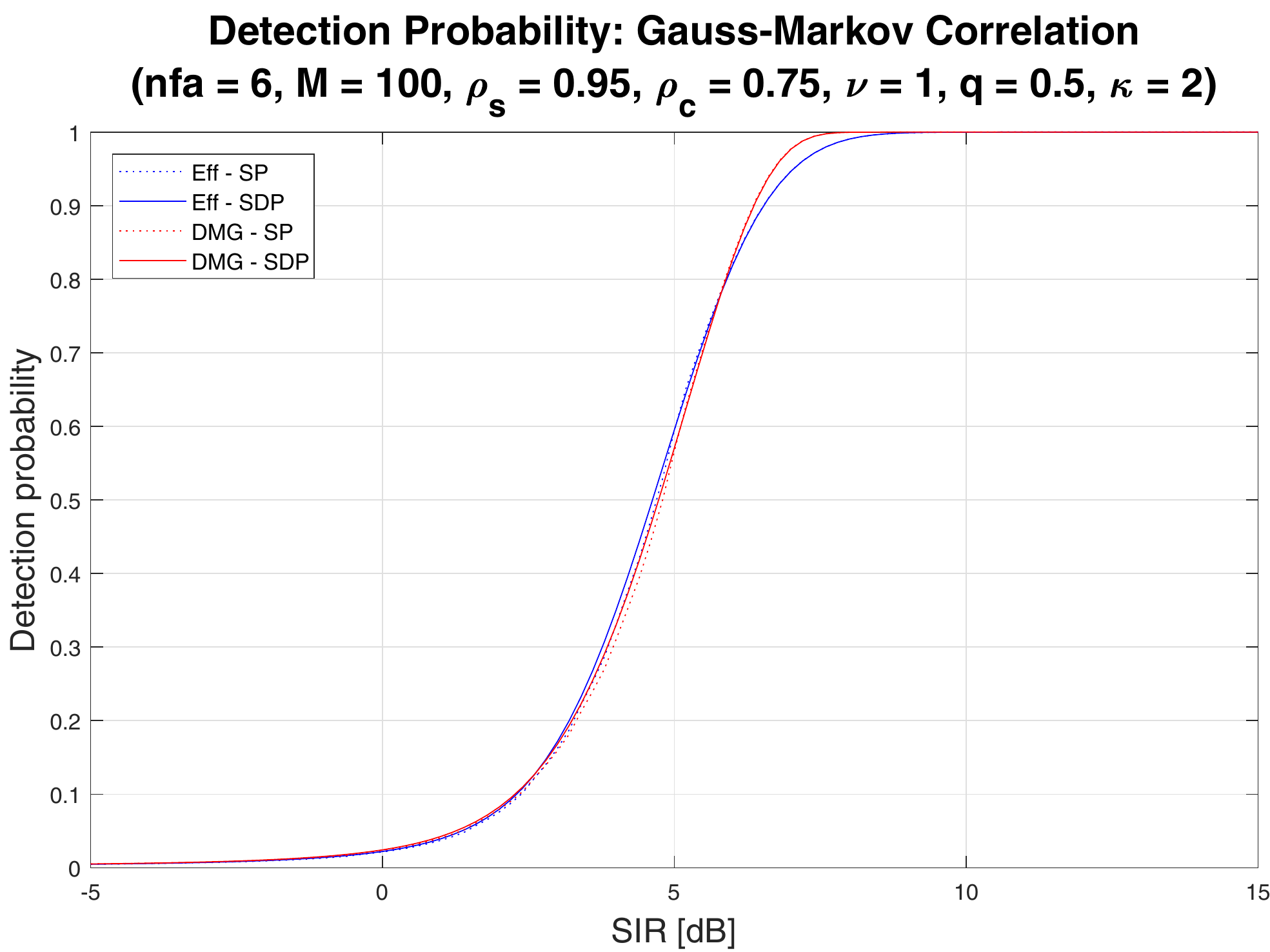}
\caption{Detection probability for a $\kappa = 2$ correlated target
  in correlated K-clutter
  according to the various computational schemes.}
\label{fig:DetGamCor}
\end{figure}

\figurename~\ref{fig:DetGamCor} shows the detection probability for a
false-alarm probability of
\mbox{$P_{\text{FA}} = 10^{-n_{\text{FA}}}$},
\mbox{$n_{\text{FA}} = 6$}
as a function of SIR for both exact effective model and DMG approximation,
each computed via integration along the SDP and with the basic SP approximation.
The difference between the SDP integration and SP approximation
(solid versus dashed lines) is barely perceptible.

\subsection{The Diagonal Approximation}
Unlike the DMG approximation of the previous section,
which has a theoretical basis arising from the
effective looks concept of dealing with correlation,
the diagonal approximation that we now introduce
is an {\it ad-hoc}\ scheme, albeit one that turns out to be remarkably accurate,
while conceptually simple, and exerting a somewhat reduced computational load
relative to the full effective model.
It comprises adoption of the effective model
while pretending that the clutter and target correlation matrices commute,
and can therefore be represented by their diagonalized forms.
Then, as with the DMG approximation, we set
\begin{align}
\begin{aligned}
a_{qm}(u) &= \left(1 - q + qu\gamma^{\text{c}}_m\right)/M \,, \\
a_m(u)    &= a_{qm}(u) + S\gamma^{\text{s}}_m/(\kappa M) \,,
\end{aligned}
\end{align}
where
\begin{align}
\begin{aligned}
\{\gamma^{\text{c}}_m: m = 1,2,\ldots,M\} &= \eig(\mtx{C}_{\text{c}}) \,, \\
\{\gamma^{\text{s}}_m: m = 1,2,\ldots,M\} &= \eig(\mtx{C}_{\text{s}}) \,,
\end{aligned}
\end{align}
but now $\mtx{C}_{\text{c}}$, $\mtx{C}_{\text{s}}$ are the actual correlation matrices
rather than their simplified DMG counterparts.

\begin{figure}[!t]
\centering
\includegraphics[width=3.45in]{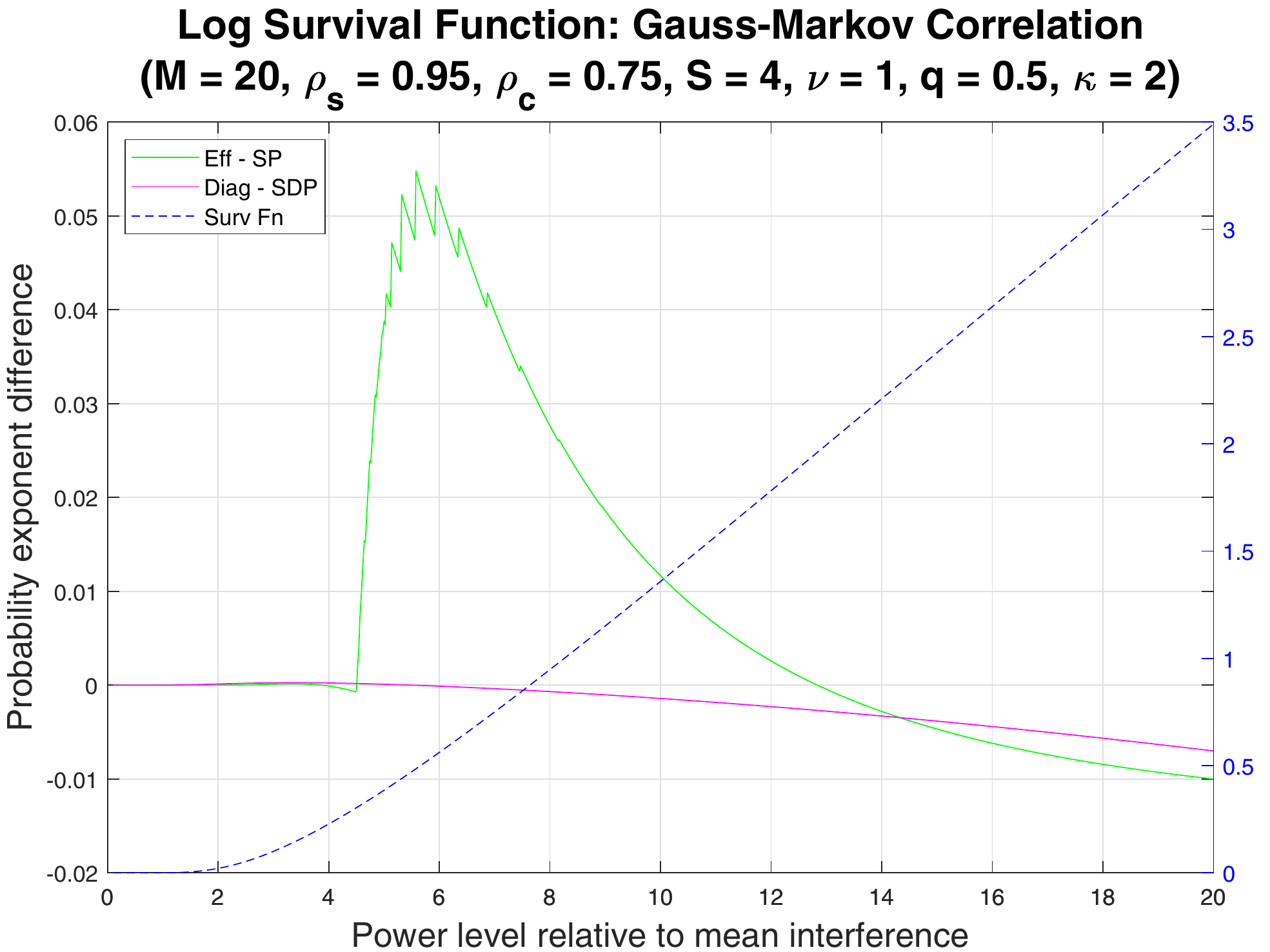}
\caption{Deviation of the survival function from the exact effective model result
     for the diagonal (magenta) and basic saddle-point (green) approximations.
     The dashed blue curve is the negative logarithm of the exact survival
     probability with scale displayed on the RHS.}
\label{fig:DiagDiff}
\end{figure}

The advantage of this over the full effective model is that, in the presence of
compound clutter, eigenvalues of a target-plus-clutter correlation matrix
do not have to be recomputed separately for each texture integration quadrature
node.
The diagonal approximation is manifestly exact for the following special cases:
\begin{itemize}
\item
when
\mbox{$\rho_{\text{c}} = 0$}
or
\mbox{$\rho_{\text{s}} = 0$};
\item
when
\mbox{$\rho_{\text{c}} = \rho_{\text{s}}$},
provided both the target and clutter have the same correlation structure;
\item
when
\mbox{$M = 2$}
and in the
\mbox{$M\to\infty$}
limit;
\item
and obviously when
\mbox{$S = 0$}
or
\mbox{$q = 0$}.
\end{itemize}
It exhibits high accuracy across the full range of parameters.
The deviation with respect to the exact result increases with power level $v$,
but lies well within 1\% for survival probabilities above $10^{-3}$.
This is illustrated in \figurename~\ref{fig:DiagDiff}
for a case where the deviation is amplified.

The magenta curve shows the difference in base-10 logarithms of the survival
function between the exact effective model and the diagonal model. The
green curve is the corresponding difference between the exact effective model
and its basic saddle-point approximation.
It can be seen that the basic SP approximation is worst around the mean of the
distribution
\mbox{$v = \bar{v}$},
which occurs at a value of the power level normalized to mean interference
given by
\mbox{$\bar{v} = 1 + S = 6$}.
The diagonal approximation slowly degrades with increasing
power level $v$, but is generally superior.
The power-axis interval shown in the graph spans a range of survival
probabilities from unity to $10^{-3.5}$, as can be seen from
the dashed blue curve representing the exact negative log-survival function.

\subsection{Effective Model for Non-Fluctuating Targets}
To derive the large-$\kappa$ limit of the effective model, we can look at
\begin{align}
\begin{aligned}
&\left[\frac{\det(1/s + \sigma^2_{\text{n}} + \sigma^2_{\text{c}}\mtx{C}_{\text{c}})}
     {\det(1/s + \sigma^2_{\text{n}} + \sigma^2_{\text{c}}\mtx{C}_{\text{c}} +
     \sigma^2_{\text{s}}\mtx{C}_{\text{s}}/\kappa)}\right]^\kappa \\
&= \frac{1}{\det(\mathbb{I} + \sigma^2_{\text{s}}\mtx{Q}^{-1}(s)\mtx{C}_{\text{s}}/\kappa)^\kappa}
\; \asym{\kappa\to\infty} \; e^{-\sigma^2_{\text{s}}\Tr\{\mtx{C}_{\text{s}}\mtx{Q}^{-1}(s)\}} \,,
\end{aligned}
\end{align}
from which we infer that
\begin{equation}
\MGF_Z(s) \asym{\kappa\to\infty} \frac{1}{s^M\det\mtx{Q}(s)}
     e^{-\sigma^2_{\text{s}}\Tr\{\mtx{C}_{\text{s}}\mtx{Q}^{-1}(s)\}} \,.
\label{LZAsym}
\end{equation}
Equivalently, including the texture variable,
\begin{equation}
\begin{split}
\MGF^{\text{spk}}_{Z'_{\text{avg}}}(s,u)& \asym{\kappa\to\infty} \exp\Biggl[-\Tr\Bigl\{
     \ln\left(\mathbb{I} + (s/M)(1-q+qu\mtx{C}_{\text{c}})\right) \\
&    {}+ (s/M)\frac{S\mtx{C}_{\text{s}}}{\mathbb{I} + (s/M)(1-q+qu\mtx{C}_{\text{c}})}
     \Bigr\}\Biggr] \,.
\end{split}
\end{equation}
We may now write
\mbox{$\mtx{C}_{\text{c}} = \mtx{R}^{-1}_{\text{c}}\mtx{\Lambda}_{\text{c}}\mtx{R}_{\text{c}}$},
so that
\begin{equation}
\begin{split}
&\Tr\frac{\mtx{C}_{\text{s}}}{\mathbb{I} + (s/M)(1-q+qu\mtx{C}_{\text{c}})} \\
&= \sum_{m=1}^M\frac{\left[\mtx{R}_{\text{c}}\mtx{C}_{\text{s}}\mtx{R}^{-1}_{\text{c}}\right]_{mm}}
     {1 + (s/M)(1-q+qu\gamma_m^{\text{c}})} \,,
\end{split}
\end{equation}
and introduce the quantities
\begin{equation}
b_m \, \equiv \, (1/M)\left[\mtx{R}_{\text{c}}\mtx{C}_{\text{s}}\mtx{R}^{-1}_{\text{c}}\right]_{mm}
     \, = \, (1/M)\sum_{n=1}^M\left[\mtx{R}_{\text{c}}\mtx{R}_{\text{s}}^{\sf T}\right]_{mn}^2
     \gamma^{\text{s}}_n \,.
\label{bcoefs}
\end{equation}
This leads to the representation
\begin{equation}
\begin{split}
&\MGF^{\text{spk}}_{Z'_{\text{avg}}}(s,u) \asym{\kappa\to\infty} {} \\
&\exp\left\{-\sum_{m=1}^M\left[
     \ln(1 + a_m(u)s) + S{\cdot}\frac{b_m s}{1 + a_m(u)s}
     \right]\right\} \,,
\label{LKapInf}
\end{split}
\end{equation}
and it follows directly from (\ref{bcoefs}) that
\mbox{$\sum_{m=1}^M b_m = 1$}
and that
\mbox{$b_m \ge 0$}
for all
\mbox{$m = 1,2,\ldots,M$}.
For a true steady target, in which case
\mbox{$\mtx{C}_{\text{s}} = \mtx{\Omega}$},
where
\mbox{$\mtx{\Omega}_{ij} \equiv 1$}
for all $i,j$,
we obtain
\begin{equation}
b_m = \frac{1}{M}\left(\sum_{n=1}^M[\mtx{R}_{\text{c}}]_{mn}\right)^2 \,.
\label{bKapInf}
\end{equation}

A Swerling-0 target in the first-principles model
(given the same target correlation matrix) yields,
in (\ref{XQ}), the expectation value
\begin{equation}
\left\langle e^{-i\mathbf{u}\cdot\mathbf{X}_{\text{s}}}\right\rangle_{\mathbf{X}_{\text{s}}}
     = e^{-i\mathbf{u}\cdot\mathbf{J}} \,,
\end{equation}
with
\mbox{$\mathbf{J} \equiv \sigma_{\text{s}}{\cdot}(1,\ldots,1)^{\sf T}$},
which leads to the MGF being given precisely by the RHS of (\ref{LZAsym}),
upon substituting
\mbox{$\mtx{C}_{\text{s}} = \mtx{\Omega}$}
therein.
Therefore, we see that the effective and first-principles models concur in the
case of a Swerling-0 target (which, by definition, is fully correlated) in correlated clutter.
In the opposite extreme, where
\mbox{$\mtx{C}_{\text{s}} = \mathbb{I}$},
we have
\begin{equation}
\left\langle e^{-i\mathbf{u}\cdot\mathbf{X}_{\text{s}}}\right\rangle_{\mathbf{X}_{\text{s}}}
     = \prod_{m=1}^M\cos\left(\sigma_{\text{s}}u_m\right) \,,
\end{equation}
which can be shown to yield the MGF
\begin{equation}
\begin{split}
\MGF_Z(s) &= \frac{1}{s^M\det\mtx{Q}(s)}e^{-\sigma^2_{\text{s}}\Tr\{\mtx{Q}^{-1}(s)\}} \\
& {}\times \prod_{\substack{i,j = 1\\i \neq j}}^M \cosh\left(\sigma^2_{\text{s}}
     \mtx{Q}^{-1}_{ij}(s)\right) \,.
\end{split}
\end{equation}
This result provides evidence that the effective and first-principles models
are not strictly identical. However, as discussed in the Appendix, the
fully phase-uncorrelated steady target is an exceptional case.
In the general partially correlated case, the result generalizes to
\begin{equation}
\begin{split}
\MGF_Z(s) &= \frac{1}{s^M\det\mtx{Q}(s)}e^{-\sigma^2_{\text{s}}\Tr\{\mtx{C}_{\text{s}}\mtx{Q}^{-1}(s)\}} \\
&    {}\times \prod_{\substack{i,j = 1\\i \neq j}}^M
     \cosh\left(\sigma^2_{\text{s}}[\mtx{L}_{\text{s}}\mtx{Q}^{-1}(s)\mtx{L}_{\text{s}}^{\sf T}]_{ij}\right) \,,
\label{FPConj}
\end{split}
\end{equation}
which follows from the representation
\mbox{$\mathbf{X}_{\text{s}} = \sigma_{\text{s}}\mtx{L}_{\text{s}}^{\sf T}\mathbf{B}$},
where the $B_m$ are iid Rademacher RVs.

\section{Monte Carlo Simulation}
\label{mcsim}
While the first-principles model, as defined in (\ref{LZFP}), seems
analytically intractable, it is highly amenable to
Monte Carlo simulation, from which an empirical survival function
can be derived.
This enables a comparison with the effective model whose survival
function can be explicitly computed via saddle-point techniques.
Therefore,
we can call upon the Kolomogorov-Smirnov statistic to test the
null hypothesis that the
empirical survival function derived from MC simulation of the
first-principles model and the analytically computed survival function
for the effective model represent the same distribution.

A Monte Carlo (MC) simulation of the first-principles model,
defined by (\ref{LZFP}),
can be easily performed at the quadrature component level.
In this case,
the returned power RV is realized according to
\begin{equation}
\begin{split}
&Z_{\text{avg}}/\sigma_{\text{I}}^2 = {} \\
&\frac{1}{2M}\sum_{m=1}^M\sum_{\eta=\text{i,q}}
     \left(\sqrt{1-q}H_{\text{n},\eta m} + \sqrt{qU}\hat{X}_{\text{c},\eta m} +
     \sqrt{S}\hat{X}_{\text{s},\eta m}\right)^2 \,.
\label{ZRV2}
\end{split}
\end{equation}
Assuming Gauss-Markov correlated clutter, we can write
\begin{equation}
\hat{X}_{\text{c},\eta m} = \rho_{\text{c}} \hat{X}_{\text{c},\eta m-1}  +
     \sqrt{1 - \rho^2_{\text{c}}}H_{\text{c},\eta m} \,,
\end{equation}
for
\mbox{$m = 2,3,\ldots,M$},
with
\mbox{$\hat{X}_{\text{c},\eta 1} = H_{\text{c},\eta 1}$}.
Here,
\mbox{$H_{\alpha,\eta m} \sim \mathcal{N}(0,1)$}
for both
\mbox{$\alpha = \text{n},\text{c}$}.
For the target RVs, we have
\begin{align}
\begin{aligned}
\hat{X}_{\text{s},\eta m} &= \sum_{m'=1}^M \hat{Y}_{\text{s},\eta m'}[\mtx{L}_{\text{s}}]_{m'm} \,, \\
\hat{Y}_{\text{s},\eta m} &= (2B_{\eta m} - 1)\sqrt{G_{\eta m}} \,,
\end{aligned}
\end{align}
where the $B_{\eta m}$ represent iid Bernoulli trials such that
\mbox{$B_{\eta m} \sim \text{Bin}(1,1/2)$},
and the $G_{\eta m}$ are iid gamma variates such that
\mbox{$G_{\eta m} \sim \Gamma(\kappa/2,2/\kappa)$}.
The RV $U$ generates the clutter texture distribution.
For K-distributed surface clutter,
\mbox{$U \sim \Gamma(\nu, 1/\nu)$}.

\begin{figure}[!t]
\centering
\includegraphics[width=3.45in]{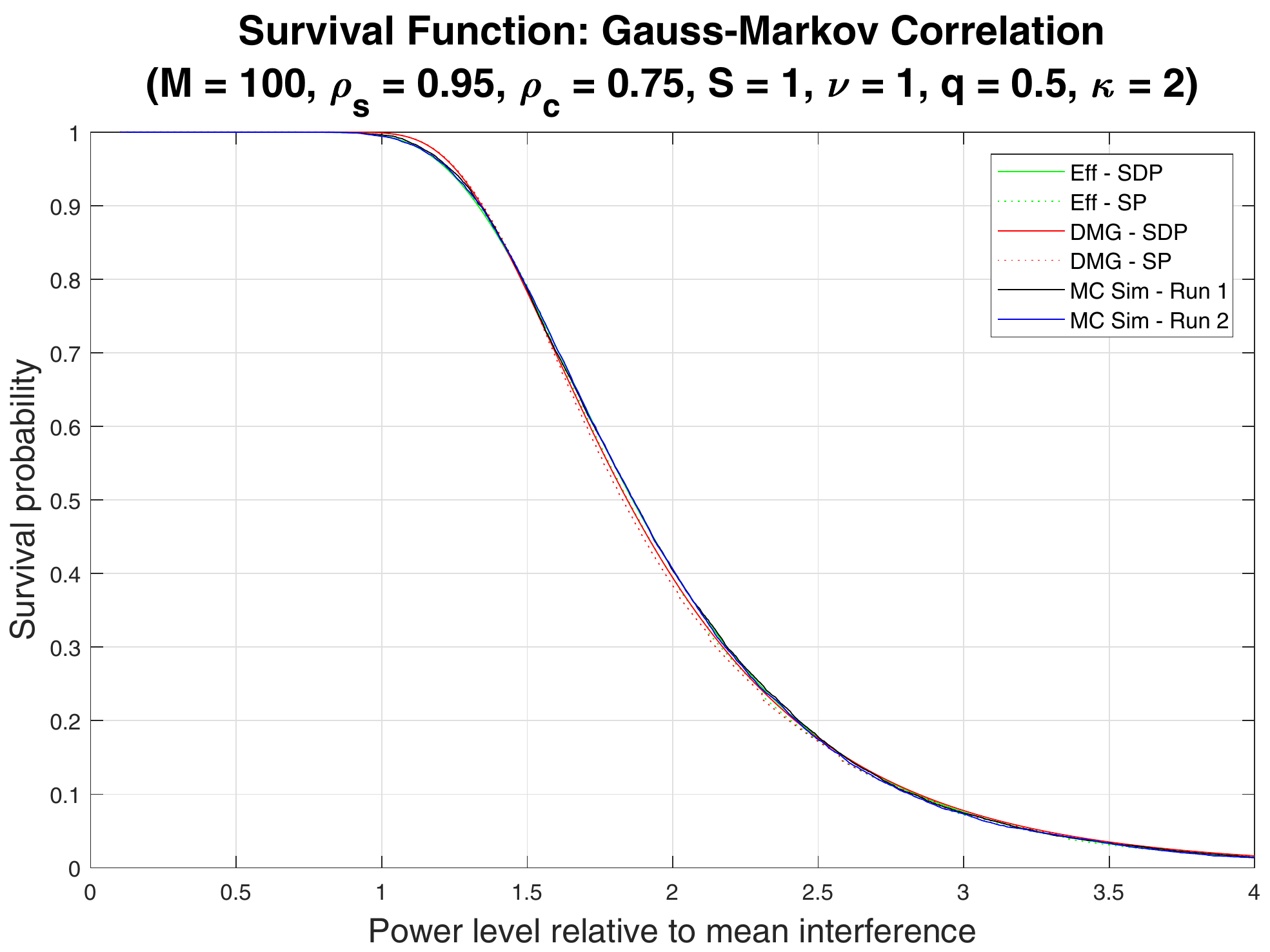}
\caption{Survival function for a $\kappa = 2$ correlated target
  in correlated K-clutter
  according to the various computational schemes.
  All curves except for those
  from the DMG approximation (red) coincide to within the line-width of the graph.}
\label{fig:SFnGamCor}
\end{figure}

\begin{figure}[!t]
\centering
\includegraphics[width=3.45in]{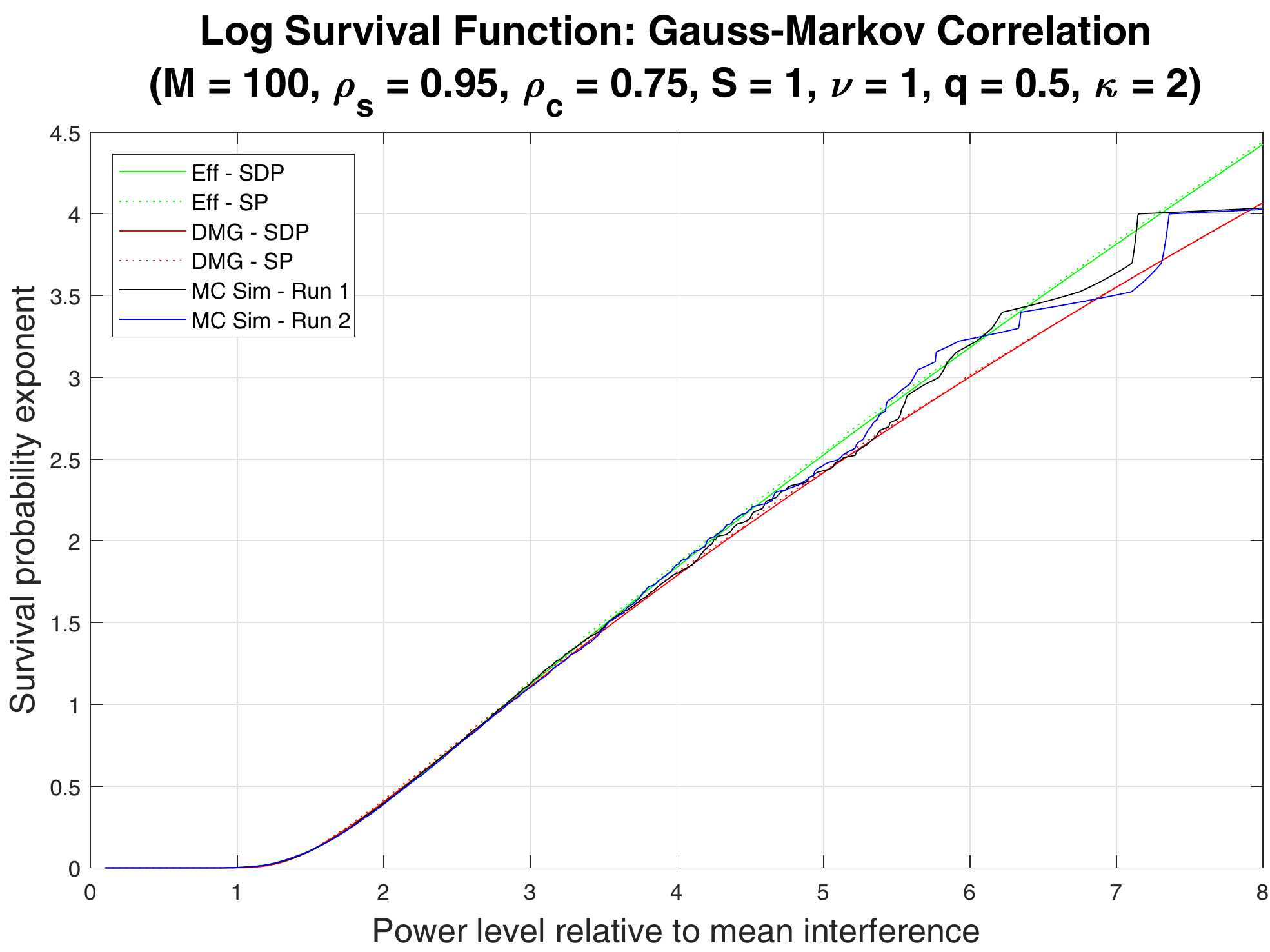}
\caption{Log-survival function for a $\kappa = 2$ correlated target
  in correlated K-clutter
  according to the various computational schemes.
  For the effective model (green) and DMG approximation (red), the dashed curve for
  the SP approximation coincides with the solid SDP curve to within the
  line-width of the graph.}
\label{fig:TailGamCor}
\end{figure}

With the aid of MC simulation, we can compare the empirical survival function
for the first-principles model with the computed survival function for the
effective model.
\figurename~\ref{fig:SFnGamCor}
compares two MC runs of the first-principles model for a gamma-fluctuating
target with
\mbox{$\kappa = 2$}
with the effective and DMG models each computed via integration along the SDP
and basic SP approximation. A high level of coincidence is observed among all the cases.
\figurename~\ref{fig:TailGamCor}
draws the same survival function on a logarithmic scale where some discrepancies become
apparent for the DMG approximation (red curve) in the tail.
The dashed green and red curves for the effective model and DMG approximation arising
from the SP approximation coincide with their SDP counterparts to within the line-width
of the graph.
Also, no difference between the first-principles (black and blue curves) and effective
(green curve) models is evident.
The MC sample size for each run was
\mbox{$N_{\text{mc}} = 10^4$}.

\begin{figure}[!t]
\centering
\includegraphics[width=3.45in]{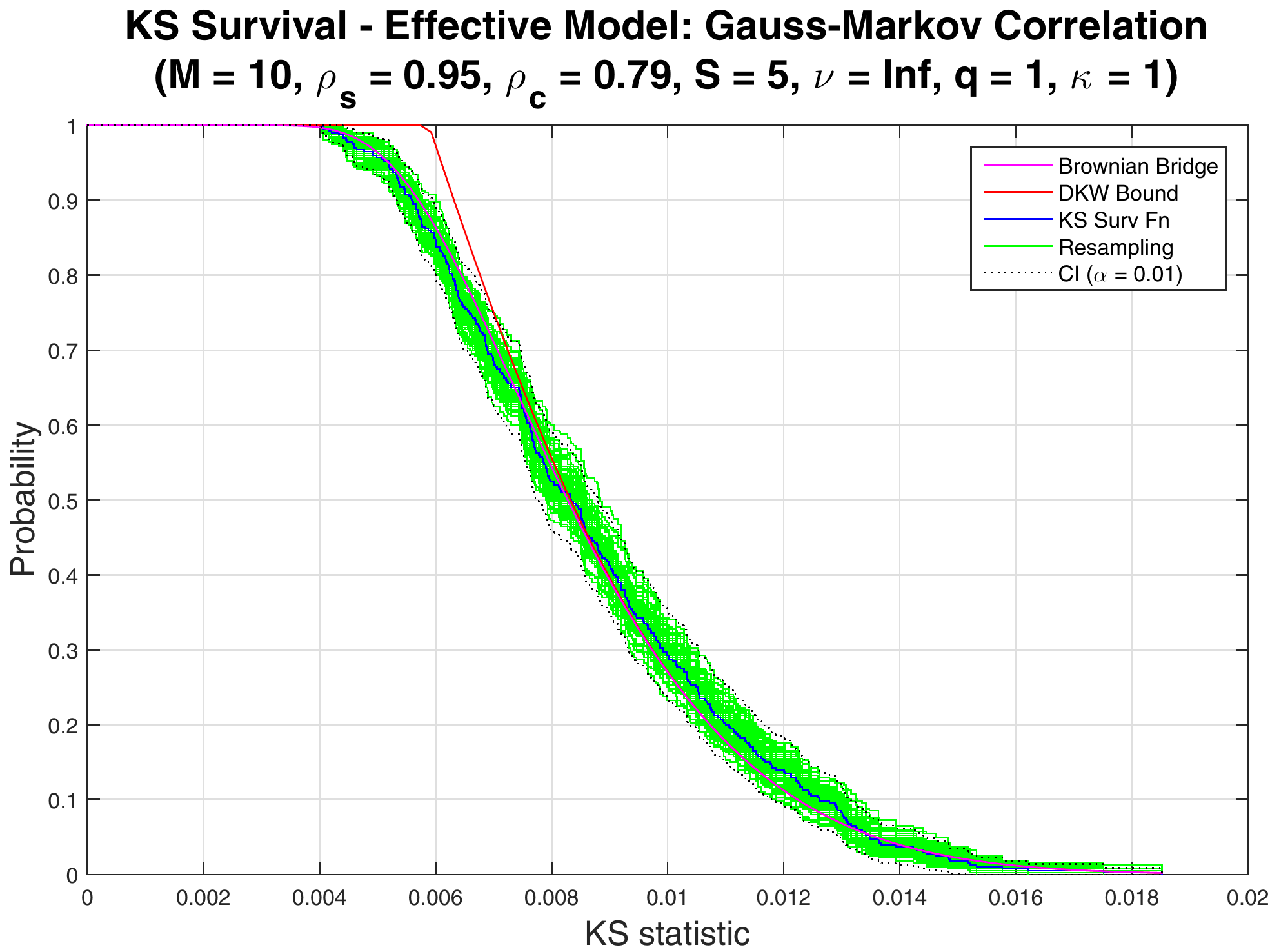}
\caption{Empirical KS survival function for a $\kappa = 1$ correlated target
  in correlated K-clutter,
  comparing the first-principles and effective models.}
\label{fig:KSSurvFnEff1}
\end{figure}

\begin{figure}[!t]
\centering
\includegraphics[width=3.45in]{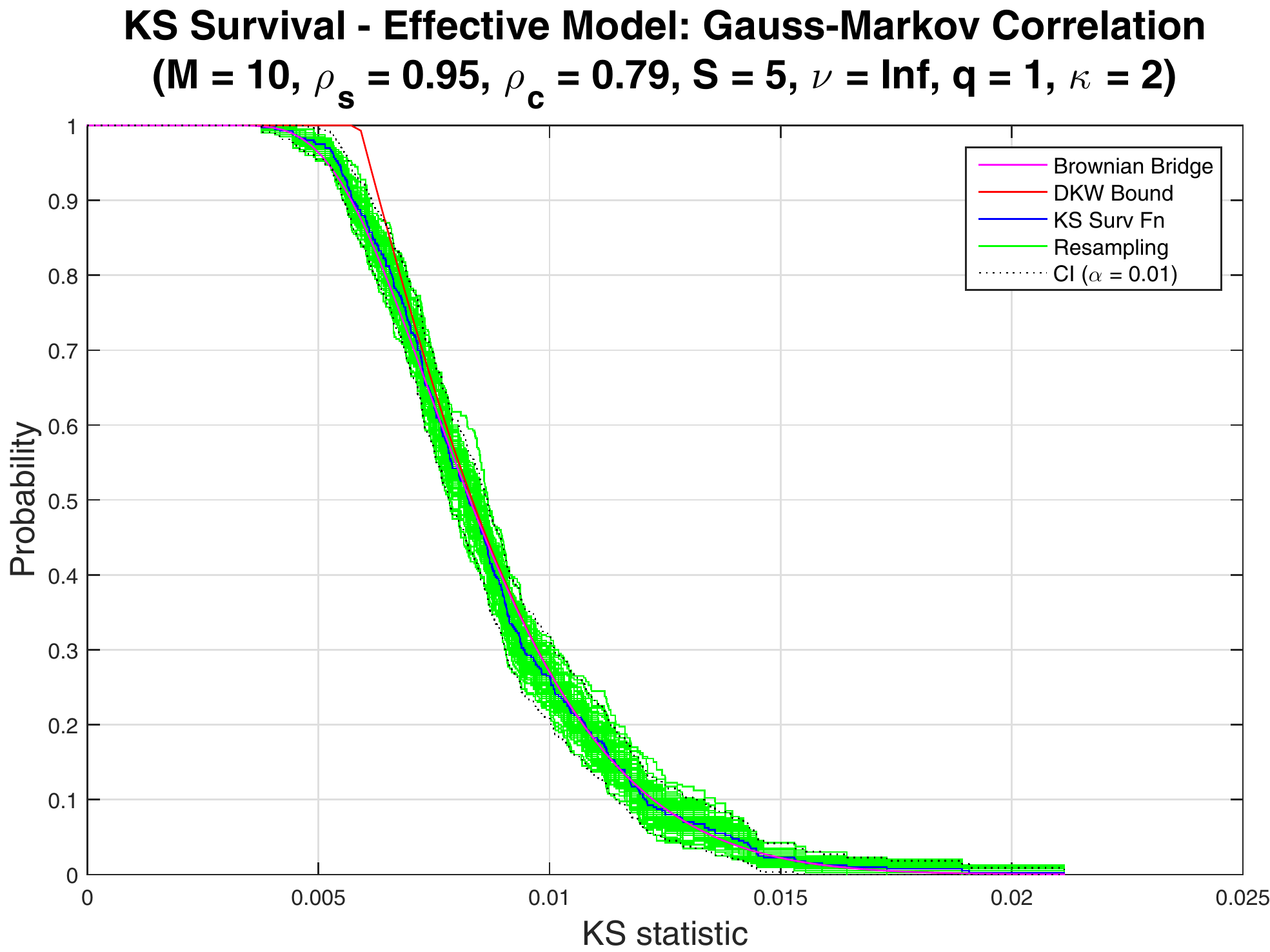}
\caption{Empirical KS survival function for a $\kappa = 2$ correlated target
  in correlated K-clutter,
  comparing the first-principles and effective models.}
\label{fig:KSSurvFnEff2}
\end{figure}

The null hypothesis that the two models generate the same
distribution can be tested by means of the one-sample Kolmogorov-Smirnov (KS)
statistic \cite{GC:Massey51}.
Since the population of returned powers associated with the first-principles
model is sampled by MC simulations, we are able generate repeated samples, and
thus construct an empirical distribution for the KS statistic.
The results for the empirical survival function of the KS statistic are presented in
\figurename~\ref{fig:KSSurvFnEff1} and \figurename~\ref{fig:KSSurvFnEff2}
for target fluctuation classes
\mbox{$\kappa = 1,2$},
respectively.

We find that we are unable to reject the null hypothesis that the first-principles
and effective models generate the same distribution at the
\mbox{$\alpha = 0.01$}
significance level for almost any combination of parameters, with the exception of
\mbox{$\kappa > 6$}
whenever
\mbox{$\mtx{C}_{\text{s}}$}
is very close (but not exactly equal) to the identity matrix,
in which case a small discrepancy is detectable.
The green region is generated by empirical survival functions derived
via bootstrap resampling, and roughly delineates the 1\% confidence interval.
The red curve is the Dvoretzky-Kiefer-Wolfowitz (DKW) bound.
The appearance of white space between the
red curve and the lower edge of the green bootstrap region indicates
that the null hypothesis should be rejected.
The magenta curve is the survival function for the Kolmogorov distribution
that represents the theoretical distribution for the KS statistic under
the null hypothesis, and is generated by the stochastic process known as the
Brownian bridge.
We note that the red DKW bound essentially tracks the Brownian
bridge due to the large sample size. We have also explicitly indicated the upper
and lower confidence limits as computed from Greenwood's formula
for a significance level of
\mbox{$\alpha = 0.01$},
indicated by the dashed black curves. These curves are seen to bound the green
bootstrap region.

\begin{figure}[!t]
\centering
\includegraphics[width=3.45in]{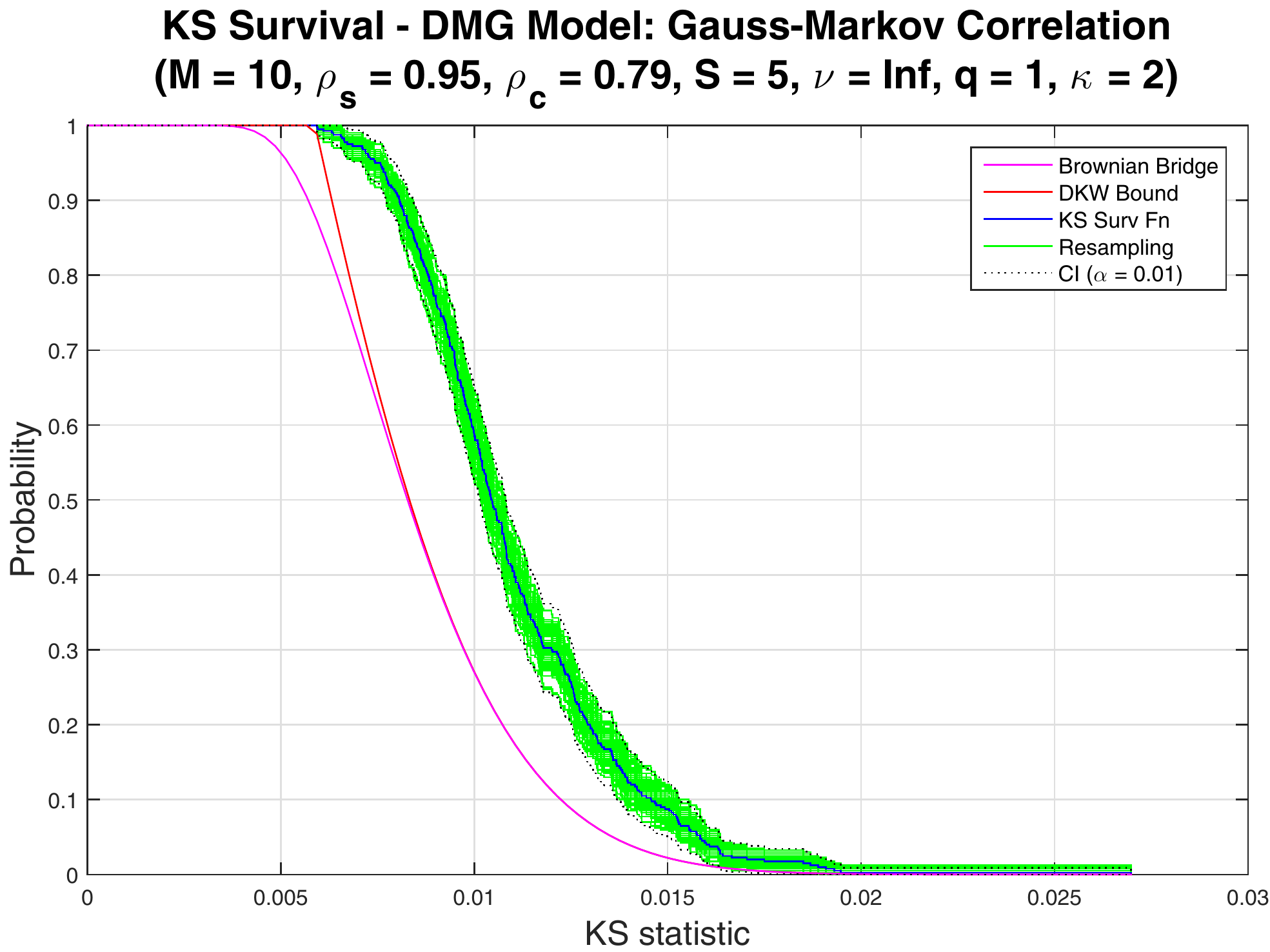}
\caption{Empirical KS survival function for a $\kappa = 2$ correlated target
  in correlated K-clutter,
  comparing the first-principles model with the DMG approximation.}
\label{fig:KSSurvFnDMG}
\end{figure}

We have also compared the first-principles model with the DMG approximation.
The results for the empirical KS survival function for a
\mbox{$\kappa = 2$}
target are presented in
\figurename~\ref{fig:KSSurvFnDMG}.
Even though the DMG approximation agrees well with both the effective and
first-principles model, as observed in the previous graphs, our statistical
test has more than sufficient power to unambiguously reject the null hypothesis
that it corresponds to the same distribution.

\subsection{Statistical Power}
\label{power}
False alarms occur in cases where the null hypothesis is rejected when it
should not have been. Missed detections occur in cases where the null
hypothesis is not rejected when it should have been.
The power of a statistical test is formally defined as one minus the
probability of a missed detection. The statistical power generally increases
with sample size, and is a measure of the sensitivity of the statistical test.

In order to quantify the power ({\it i.e.}\ sensitivity) of the KS test
employed in the foregoing section, one may ask how much must one deform
the distribution of the effective model away from its true functional
form before the null hypothesis that it agrees with the empirical
distribution of the first-principles model is rejected at the desired
confidence level.
In this section, we show that the required amount of deformation is tiny.
In other words, if the survival function of the effective model were only
very slightly different, then a discrepancy with respect to the
first-principles model would be detected.

\begin{figure}[!t]
\centering
\includegraphics[width=3.45in]{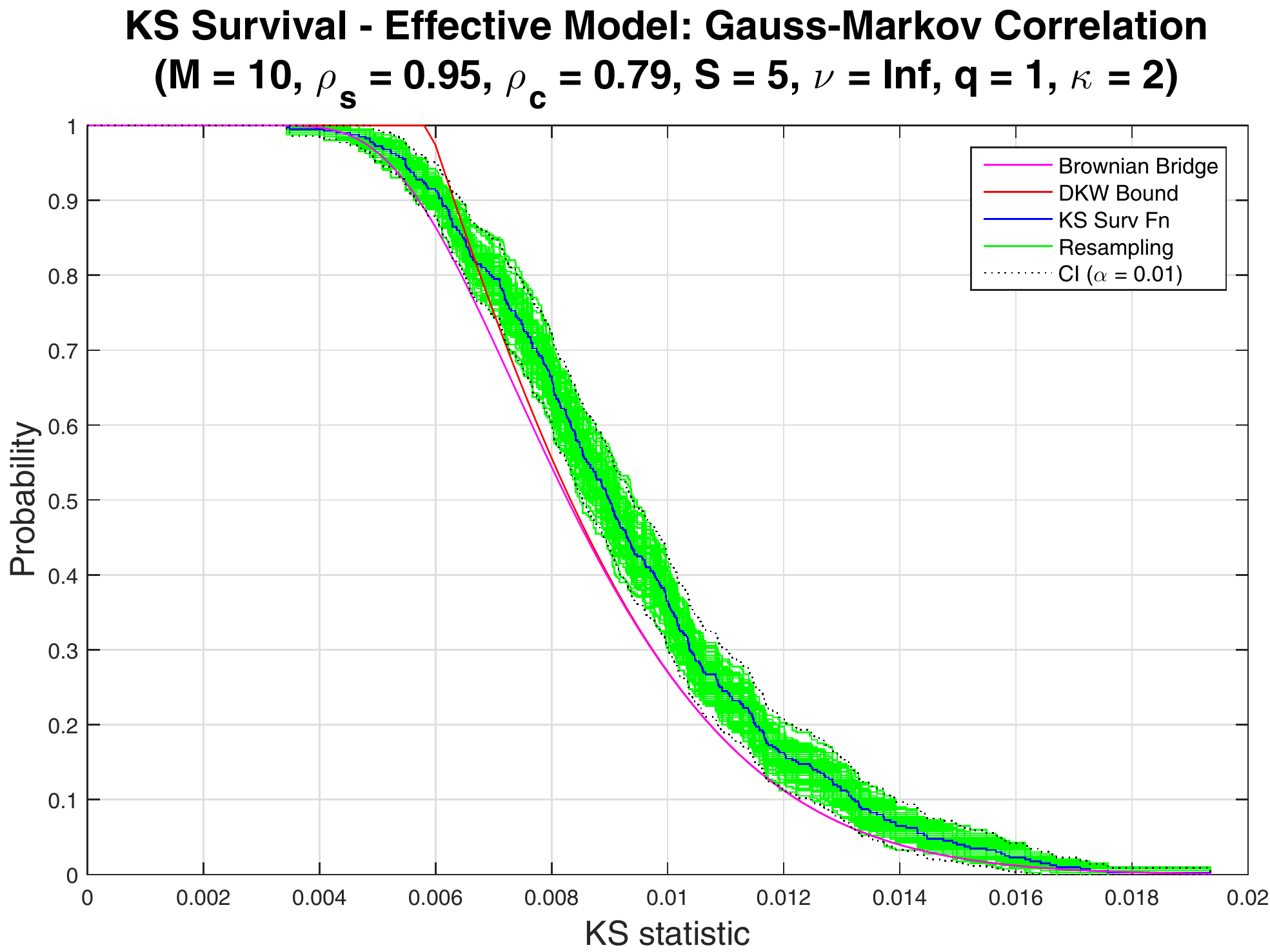}
\caption{Comparison of the empirical survival function relative to the perturbed effective
        model with $\delta_{\text{max}} = 0.003$, showing the $\alpha = 0.01$ confidence
        interval, and the theoretical curve under the null hypothesis.}
\label{fig:KSPert}
\end{figure}

\begin{figure}[!t]
\centering
\includegraphics[width=3.45in]{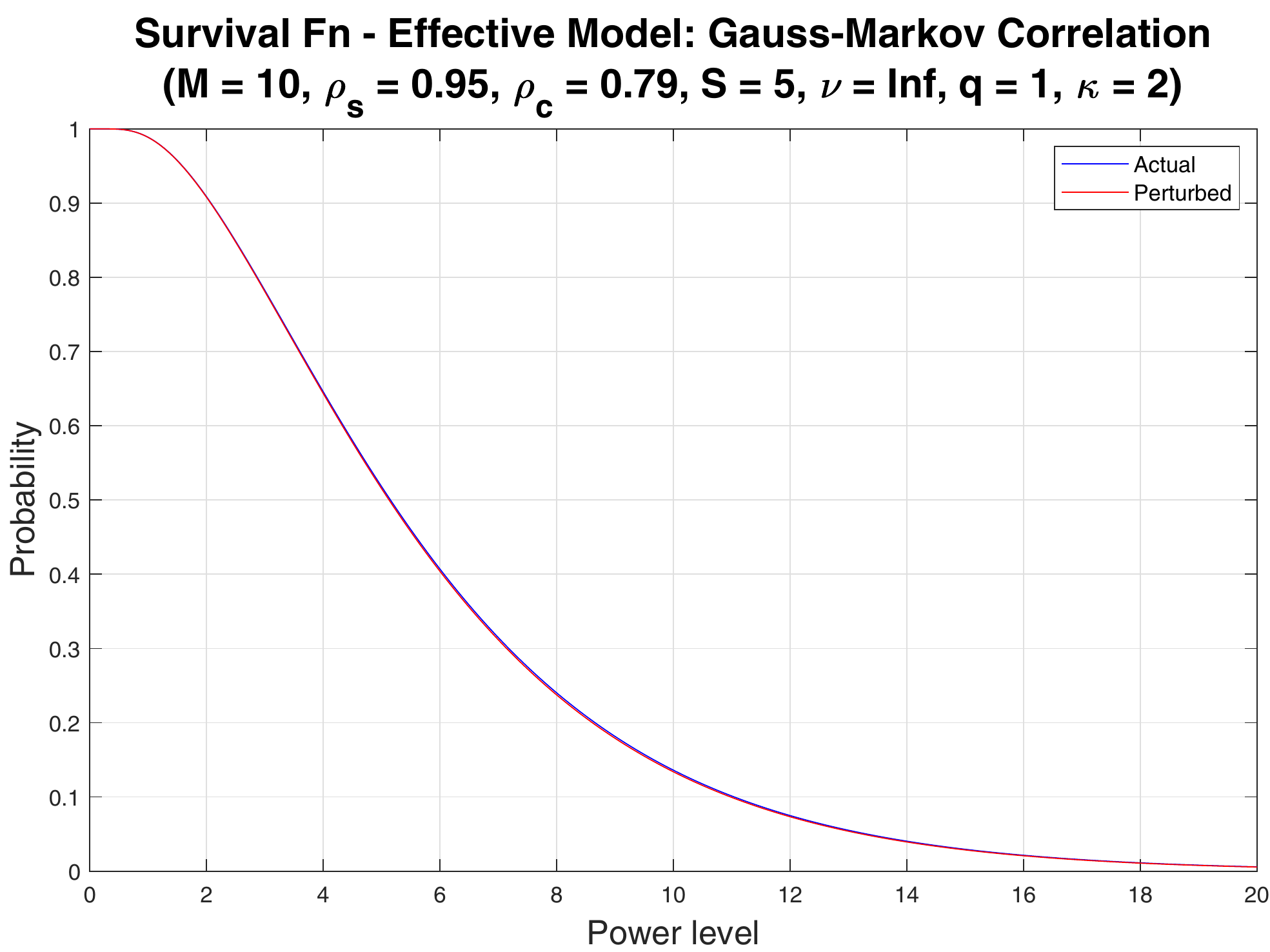}
\caption{Comparison of the survival function for the effective model with its
         perturbed counterpart for $\delta_{\text{max}} = 0.003$ with
         the same parameters as in the previous figure.}
\label{fig:SFnComp}
\end{figure}

Thus, in order to characterize the statistical power of the KS test, we shall
consider, as an alternative hypothesis, the slightly perturbed effective-model
survival function \cite{GC:Suzuki68,GC:Schultz72}
\begin{equation}
G(x) \equiv \left[\bar{F}(x)\right]^{1+\epsilon} \,,
\end{equation}
where
\mbox{$|\epsilon| \ll 1$}.
The value of the perturbation parameter $\epsilon$ can be related to the
maximum difference $\delta_{\text{max}}$ between $\bar{F}(x)$ and $G(x)$.
Writing
\begin{equation}
\delta(x) \equiv \bar{F}(x) - \left[\bar{F}(x)\right]^{1+\epsilon} \,,
\end{equation}
we obtain the derivative
\begin{equation}
\delta'(x) = f(x)\left[(1+\epsilon)\bar{F}(x)^\epsilon - 1\right] \,.
\end{equation}
where
\mbox{$f(x)$}
denotes the PDF associated with
\mbox{$\bar{F}(x)$}.
The extremum condition
\mbox{$\delta'(x) = 0$}
is satisfied when
\begin{equation}
\bar{F}(x) = \left(\frac{1}{1+\epsilon}\right)^{1/\epsilon} \,.
\end{equation}
Hence, we find that
\begin{equation}
\delta_{\text{max}} \; = \;  \bar{F}(x)\left[1 - \bar{F}(x)^\epsilon\right]
\; = \; \epsilon{\cdot}\left(\frac{1}{1+\epsilon}\right)^{1+1/\epsilon} \,.
\end{equation}
This may be expressed as
\begin{align}
\begin{aligned}
\ln\delta_{\text{max}} &= \ln\epsilon - (1+1/\epsilon)\ln(1+\epsilon) \\
&= \ln\epsilon - 1 +O(\epsilon^2) \,,
\end{aligned}
\end{align}
in which case we obtain the approximate inversion
\mbox{$\epsilon \simeq e{\cdot}\delta_{\text{max}}$}.
We may now choose small values of $\delta_{\text{max}}$ and study
at which point the theoretical distribution of the KS statistic
under the null hypothesis falls
outside the confidence intervals of the empirical distribution.
With
\mbox{$N_{\text{mc}} = 10^4$}
MC trials, a KS sample size of
\mbox{$K = 400$},
and model parameters as given in \figurename~\ref{fig:KSPert},
we find a statistical power close to unity at a significance level of
\mbox{$\alpha = 0.01$}
when
\mbox{$\delta_{\text{max}} = 0.003$}.
The null and alternative survival functions for this value of $\delta_{\text{max}}$
are plotted in \figurename~\ref{fig:SFnComp},
where they are seen to be barely distinguishable.

It is clear in \figurename~\ref{fig:KSPert},
that the Brownian bridge, represented by the magenta curve, lies just outside the
\mbox{$\alpha = 0.01$}
confidence intervals,
indicated by the dashed black curves,
and hence the situation depicted here shows the smallest
deformation of the effective-model survival function that leads to a rejection
of the null hypothesis.

The upshot of this is that while the KS test is able to detect the difference
between the effective model and a deformation of it that is barely discernible,
as shown in \figurename~\ref{fig:SFnComp},
it is unable to detect a discrepancy between the effective and first-principles
models at the same significance level. It follows that the effective model is a
sound proxy for the first-principles model.

\section{Discussion}
\label{Results}
The first-principles and effective models always exhibit the same mean and variance,
and are in exact agreement in the following situations:
\begin{itemize}
\item
Uncorrelated clutter
(\mbox{$\mtx{C}_{\text{c}} = \mathbb{I}$});
\item
Gaussian target
(\mbox{$\kappa = 1$})
with arbitrary target and clutter correlations;
\item
Swerling-0 target
(\mbox{$\kappa\to\infty, \mtx{C}_{\text{s}} =  \Omega$})
with arbitrary clutter correlation;
\item
Fully correlated target
(\mbox{$\mtx{C}_{\text{s}} =  \mtx{\Omega}$})
with arbitrary clutter correlation;
\item
\mbox{$M \le 2$}
and
\mbox{$M\to\infty$};
\item
Commuting target and clutter correlation matrices (provided
\mbox{$\mtx{C}_{\text{s}} \neq \mathbb{I}$});
\item
Uncorrelated target
(\mbox{$\mtx{C}_{\text{s}} = \mathbb{I}$}),
(provided one sets
\mbox{$\mtx{R}_{\text{s}} = \mtx{R}_{\text{c}}$}
when
\mbox{$\kappa > 1$});
\item
And, trivially, as
\mbox{$S\to 0$}
and
\mbox{$S\to\infty$}.
\end{itemize}
The first and final five cases apply to targets in all
finite Swerling fluctuation classes
\mbox{$\kappa = 1,2,\ldots$}.

In all other cases studied, provided
\mbox{$\kappa \le 6$}
when
\mbox{$\mtx{C}_{\text{s}}$}
is close to the identity matrix,
the first-principles and effective models are seen to agree
for all practical purposes, as demonstrated by the examining the KS statistic.
Thus, the effective model is a good proxy for the first-principles model,
since the former lends itself to efficient computation via the saddle-point
technique, whereas for the latter one must resort to MC simulation.

By studying the
\mbox{$\kappa\to\infty$}
limit, we are able to show that the first-principles and effective models
do not always coincide.
The worst-case scenario occurs when
\mbox{$\rho_{\text{c}} = 1$},
\mbox{$\rho_{\text{s}} = 0^+$}
and
\mbox{$q = 1$},
with the SIR and integrated pulses around
\mbox{$S = 5$}
and
\mbox{$M = 10$},
respectively.
Physically, this represents a highly unlikely scenario.
The fluctuation parameter $\kappa$ must exceed
\mbox{$\kappa = 6$}
before we are able to reject the null hypothesis.

\begin{figure}[!t]
\centering
\includegraphics[width=3.45in]{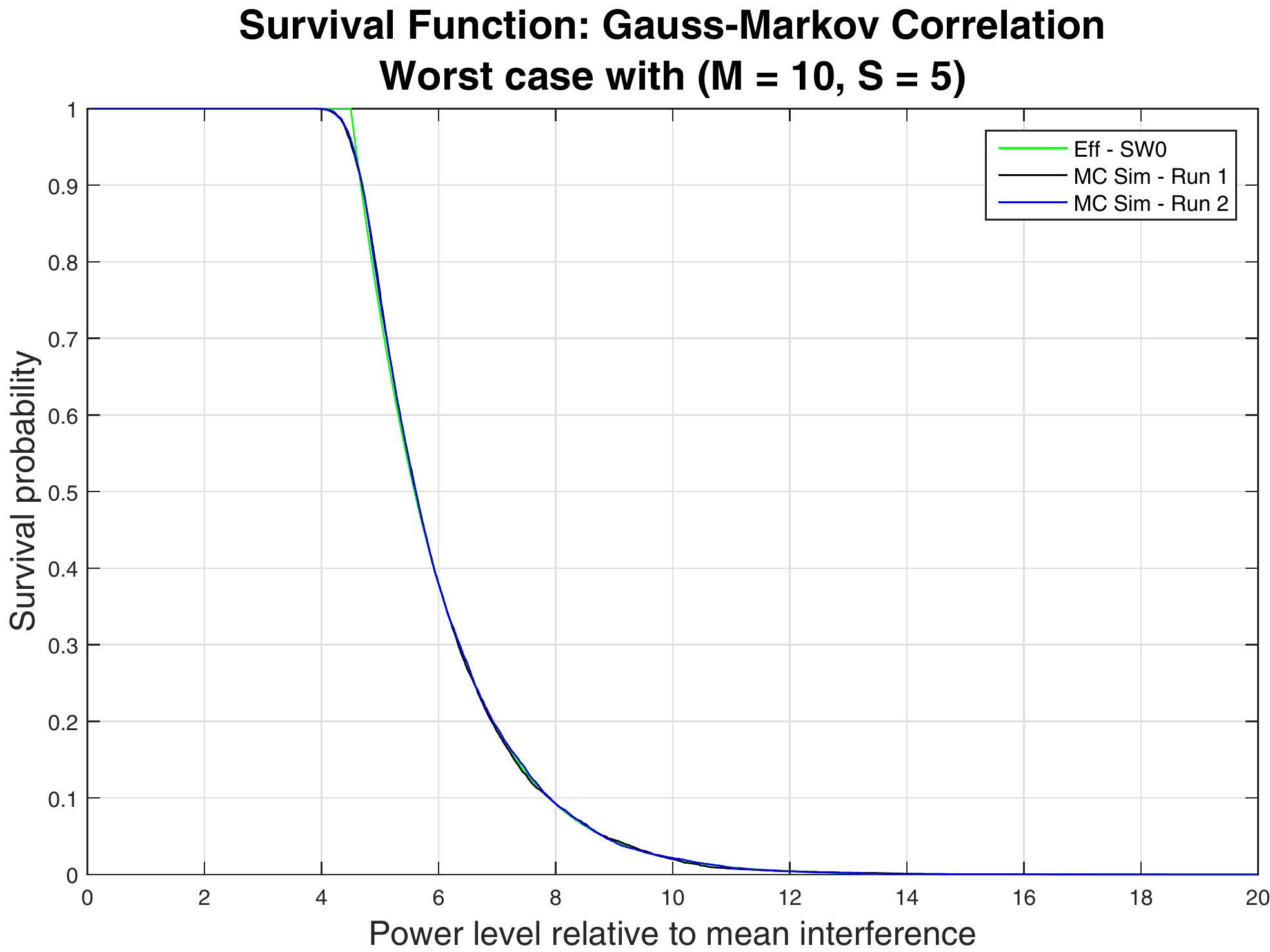}
\caption{Comparison of the survival functions for the effective model and two
         MC runs of the first-principles model in the worst case scenario.}
\label{fig:WorstCase}
\end{figure}

In order to derive the first-principles MGF for the worst case discrepancy, we set
\mbox{$q = 1$},
\mbox{$\mtx{C}_{\text{s}} = \mathbb{I}$}
and
\begin{equation}
\mtx{C}_{\text{c}} \; = \; \mtx{\Omega}  \; = \; \mtx{R}_{\text{c}}^{\sf T}\mtx{\Lambda}_{\text{c}}\mtx{R}_{\text{c}} \,,
     \quad  \mtx{\Lambda}_{\text{c}} = \diag(M,0,\ldots,0) \,,
\end{equation}
so that
\begin{equation}
\begin{split}
\MGF_{Z'_{\text{avg}}}(s)& = \frac{1}{\det[s\hat{\mtx{Q}}(s)]}e^{-(S/M)\Tr\{\hat{\mtx{Q}}^{-1}(s)\}}{\cdot} \\
& {}\times \prod_{\substack{i,j = 1\\i \neq j}}^M \cosh\left((S/M)
     \hat{\mtx{Q}}^{-1}_{ij}(s)\right) \,,
\end{split}
\end{equation}
with
\mbox{$\hat{\mtx{Q}}(s) = 1/s + \mtx{\Omega}/M$}.
Then, we have
\begin{align}
\begin{aligned}
\hat{\mtx{Q}}^{-1}(s) &= \mtx{R}_{\text{c}}^{\sf T}\diag(s/(1+s),s,\ldots,s)\mtx{R}_{\text{c}} \\
&= s\mathbb{I} - \frac{s^2}{1+s}\mtx{\Omega}/M \,,
\end{aligned}
\end{align}
in which case
\begin{equation}
\Tr\{\hat{\mtx{Q}}^{-1}(s)\} = (M-1)s + s/(1+s) \,,
\end{equation}
and so
\begin{equation}
\begin{split}
\MGF_{Z'_{\text{avg}}}(s)& = \frac{1}{1+s}e^{-(1-1/M)Ss}{\cdot}e^{-(S/M)s/(1+s)}{\cdot} \\
& {}\times \cosh^{M(M-1)}\left(\frac{(S/M^2)s^2}{1+s}\right) \,.
\end{split}
\end{equation}
On comparing this with (\ref{LZAsym}) we see that, by neglecting the
cosh term, the survival function for the corresponding effective model
is recovered and it reduces to
\begin{equation}
\bar{F}_{\text{eff}}(v;S,M) = \bar{F}_{\text{SW0}}(v - (1-1/M)S;S/M,1) \,,
\label{EffSW0}
\end{equation}
where
\mbox{$\bar{F}_{\text{SW0}}(v;S,M)$}
denotes the $M$-pulse Swerling-0 ({\it i.e.}\ Rician)
survival function for SIR $S$,
normalized such that
\begin{equation}
\int_0^\infty dv\, \bar{F}_{\text{SW0}}(v;S,M) = 1+S  \,.
\end{equation}
This serves to quantify the extent of the discrepancy.
The mean KS statistic in this worst case is around $0.04$ with the
discrepancy highly localized in the vicinity of the knee of the
survival function at
\mbox{$v = (1-1/M)S$}
as shown in \figurename~\ref{fig:WorstCase}.
The survival function for the effective model was computed from
(\ref{EffSW0})
which holds strictly in the limit
\mbox{$\rho_{\text{s}}\to 0$}.
If we select a tiny but finite value, such as
\mbox{$\rho_{\text{s}} = 10^{-4}$},
the KS statistic drops to $0.008$,
for which the null hypothesis cannot be rejected.

\begin{table}[!t]
\renewcommand{\arraystretch}{1.3}
\caption{Run-times and accuracies}
\centering
\label{tab:runtimes}
\begin{tabular}{|l||c|c|c|}
\hline
Method          & Run-time & \multicolumn{2}{|c|}{Probability error} \\
                & (sec)    & Absolute & Relative \\
\hline\hline
Effective -- SDP & $5.59$  &                        &                       \\
Effective -- SP  & $1.45$  & $3.58 \times 10^{-2}$  & $2.39 \times 10^{-3}$ \\
DMG -- SDP       & $0.94$  & $3.43 \times 10^{-2}$  & $8.45 \times 10^{-4}$ \\
DMG --  SP       & $0.66$  & $6.75 \times 10^{-2}$  & $2.30 \times 10^{-3}$ \\
Diagonal -- SDP  & $5.13$  & $1.92 \times 10^{-4}$  & $5.50 \times 10^{-5}$ \\
Diagonal --  SP  & $1.12$  & $3.58 \times 10^{-2}$  & $2.44 \times 10^{-3}$ \\
\hline
\end{tabular}
\end{table}

In Table~\ref{tab:runtimes}, we compare the run-time and accuracy of each of the methods
for calculating the effective model: These comprise the full effective model, the DMG approximation
and the diagonal approximation, each computed within the basic saddle-point approximation (SP) and
with integration along the path of steepest descent (SDP).
The run-time is that required to compute a grid of 100 points on the survival function from the
origin to the point where the probability has decreased to $10^{-3}$.
Each probability error measures the maximum deviation of the survival function
relative to the SDP result for the full effective model, expressed as both absolute and relative error.
Each run-time and error represent an average of 100 randomized parameter combinations.
The SIR is drawn in the range
\mbox{$1 \le S \le 10$},
the shape parameter is drawn in the range
\mbox{$1 \le \nu \le 10$},
and the clutter-to-interference ratio in the range
\mbox{$0.5 \le q \le 1$}.
The remaining parameters are kept constant with values
\mbox{$M = 100$},
\mbox{$\rho_{\text{c}} = 0.75$},
\mbox{$\rho_{\text{s}} = 0.95$},
\mbox{$\kappa = 2$}.
The calculations were performed in {\sc Matlab} on a laptop computer with an
Intel Core i5-4210U processor running at 1.70~GHz.
We see that the DMG approximation is very fast but is less accurate than the diagonal approximation,
which maintains a higher degree of accuracy but runs only marginally faster than the full effective model.
The main draw-card of the diagonal model is its conceptual simplicity.

\section{Conclusions}
Our main conclusion is that, for all practical purposes, the effective model,
defined in terms of its MGF, solves the problem of non-coherent detection of
a gamma-fluctuating target in the presence of
simultaneous arbitrary correlation of target
RCS and clutter speckle, and solves it in a computationally efficient manner.
One open question that arises from this work is whether there exists
an alternative first-principles model whose MGF is exactly that of the
effective model.

\section*{Acknowledgment}
I would like to thank Stephen Bocquet and Luke Rosenberg for useful discussions
and encouragement during the course of this research.

\appendix
A shortcoming of the first-principles model appears when dealing with an
uncorrelated target, for which
\mbox{$\mtx{C}_{\text{s}} = \mathbb{I}$}.
Here,
\mbox{$\mtx{\Lambda}_{\text{s}} = \mathbb{I}$},
and so
\mbox{$\mtx{L}_{\text{s}} = \mtx{R}_{\text{s}}$}
can be an arbitrary rotation matrix.
However, the MGF given in Section~\ref{FirstPrinc} is not independent of $\mtx{R}_{\text{s}}$.
A practical work-around is to select
\mbox{$\mtx{R}_{\text{s}} = \mtx{R}_{\text{c}}$}
in this case. This also serves to bring it into agreement with the
associated effective model, introduced in Section~\ref{EffModel},
which is explicitly independent of $\mtx{R}_{\text{s}}$
for an uncorrelated target.

It may also appear that the first-principles model is not well-defined
since the matrix
\mbox{$\mtx{L}_{\text{s}} = \sqrt{\mtx{\Lambda}_{\text{s}}}\mtx{R}_{\text{s}}$}
is not uniquely constructed from the eigenvalue decomposition
\mbox{$\mtx{C}_{\text{s}} = \mtx{R}_{\text{s}}^{\sf T}\mtx{\Lambda}_{\text{s}}\mtx{R}_{\text{s}}$}.
While this is true for
\mbox{$\mtx{C}_{\text{s}} = \mathbb{I}$},
it is not generally so:
The uncorrelated target is an exceptional case.
The underlying theory assumes a non-degenerate eigenvalue spectrum
(which is violated for an uncorrelated target).
The eigenvalue decomposition partitions the Euclidean space $\mathbb{R}^M$
into $M$ orthogonal directions, and the columns of the matrix $\mtx{R}_{\text{s}}$
are the respective unit vectors associated with these directions.
Each of these is uniquely
determined up to an overall sign. Thus, we also have
\mbox{$\mtx{C}_{\text{s}} = \mtx{R}_{\text{s}}^{\prime{\sf T}}\mtx{\Lambda}_{\text{s}}\mtx{R}'_{\text{s}}$}
where
\mbox{$\mtx{R}'_{\text{s}} = \mtx{B}\mtx{R}_{\text{s}}$}
for any diagonal matrix $\mtx{B}$ with diagonal elements $\pm 1$.
Fortunately, the first-principles model as given in (\ref{LZFP})
is invariant under the transformation
\mbox{$\mtx{L}_{\text{s}} \mapsto \mtx{B}\mtx{L}_{\text{s}}$}.

It is worth pointing out that the first-principles MGF can be evaluated
explicitly in the case of a fully correlated target
(\mbox{$\mtx{C}_{\text{s}} = \mtx{\Omega}$},
where
\mbox{$\mtx{\Omega}_{ij} \equiv 1$}
for all $i,j$)
for any $\kappa$.
Here, the matrix $\mtx{L}_{\text{s}}$ is such that, for any matrix $\mtx{A}$,
\begin{equation}
\mtx{L}_{\text{s}}\mtx{A}\mtx{L}_{\text{s}}^{\sf T} = \diag\left(\Tr\{\mtx{\Omega}\mtx{A}\}, 0, \ldots, 0\right) \,.
\end{equation}
In particular,
\begin{equation}
\mtx{L}_{\text{s}}\mtx{Q}^{-1}(s)\mtx{L}_{\text{s}}^{\sf T} =
     \diag\left(\Tr\{\mtx{C}_{\text{s}}\mtx{Q}^{-1}(s)\}, 0, \ldots, 0\right) \,,
\end{equation}
in which case
\begin{equation}
\hat{\mathbf{Y}}^{\sf T}(\mtx{L}_{\text{s}}\mtx{Q}^{-1}(s)\mtx{L}_{\text{s}}^{\sf T})\hat{\mathbf{Y}}
     = \Tr\{\mtx{C}_{\text{s}}\mtx{Q}^{-1}(s)\}{\cdot}\hat{Y}_1^2 \,.
\end{equation}
It follows that
\begin{equation}
\begin{split}
\MGF_Z(s) &= \frac{1}{s^M\det\mtx{Q}(s)} \\
& {}\times \left\langle\exp\left(-\frac{\sigma^2_{\text{s}}}{2}
     \Tr\{\mtx{C}_{\text{s}}\mtx{Q}^{-1}(s)\}{\cdot}\hat{Y}_1^2\right)
     \right\rangle^2_{\hat{Y}_1} \,.
\end{split}
\end{equation}
The expectation is given by
\begin{align}
\begin{aligned}
&\left\langle\exp\left(-\frac{\sigma^2_{\text{s}}}{2}
     \Tr\{\mtx{C}_{\text{s}}\mtx{Q}^{-1}(s)\}{\cdot}\hat{Y}_1^2\right)
     \right\rangle_{\hat{Y}_1} \\
&\quad =   \frac{1}{\Gamma(\kappa/2)}\int_0^\infty dx\, x^{\kappa/2-1} \\
&\quad\quad {}\times \exp\left\{-\left[1 + (\sigma^2_{\text{s}}/\kappa)
     \Tr\{\mtx{C}_{\text{s}}\mtx{Q}^{-1}(s)\}\right]x\right\} \\
&\quad = \left[1 + (\sigma^2_{\text{s}}/\kappa)
     \Tr\{\mtx{C}_{\text{s}}\mtx{Q}^{-1}(s)\}\right]^{-\kappa/2} \,.
\end{aligned}
\end{align}
Next, let us look at
\begin{align}
\begin{aligned}
&\ln\det\left[\mathbb{I} +  (\sigma^2_{\text{s}}/\kappa)\mtx{C}_{\text{s}}\mtx{Q}^{-1}(s)\right] \\
&\quad = \Tr\left\{\ln\left[\mathbb{I} +
     (\sigma^2_{\text{s}}/\kappa)\mtx{C}_{\text{s}}\mtx{Q}^{-1}(s)\right]\right\} \\
&\quad = \sum_{n=1}^\infty\frac{1}{n}\left(\sigma^2_{\text{s}}/\kappa\right)^n
     \Tr\left\{\mtx{C}_{\text{s}}\mtx{Q}^{-1}(s)\right\}^n \,,
\end{aligned}
\end{align}
and observe that
\begin{equation}
\Tr\left\{\mtx{C}_{\text{s}}\mtx{Q}^{-1}\right\}^n \; = \;
     \Tr\left\{\mtx{L}_{\text{s}}\mtx{Q}^{-1}\mtx{L}_{\text{s}}^{\sf T}\right\}^n
     \; = \; \left(\Tr\{\mtx{C}_{\text{s}}\mtx{Q}^{-1}\}\right)^n \,,
\end{equation}
from which it follows that
\begin{equation}
\MGF_Z(s) = \frac{1}{s^M\det\mtx{Q}(s)}{\cdot}\frac{1}{\left[\det\left(\mathbb{I}
     + \sigma^2_{\text{s}}\mtx{C}_{\text{s}}\mtx{Q}^{-1}(s)/\kappa\right)\right]^\kappa} \,,
\end{equation}
which, as one can see by appealing to (\ref{LZEff}) or (\ref{LZAsym}),
is in agreement with the effective model discussed
in Section~\ref{EffModel}.

\bibliographystyle{./IEEEtran}
\bibliography{./IEEEabrv,./GamCorrIEEE}
%
%
\begin{IEEEbiography}[{\includegraphics[width=1in,height=1.25in,clip,keepaspectratio]{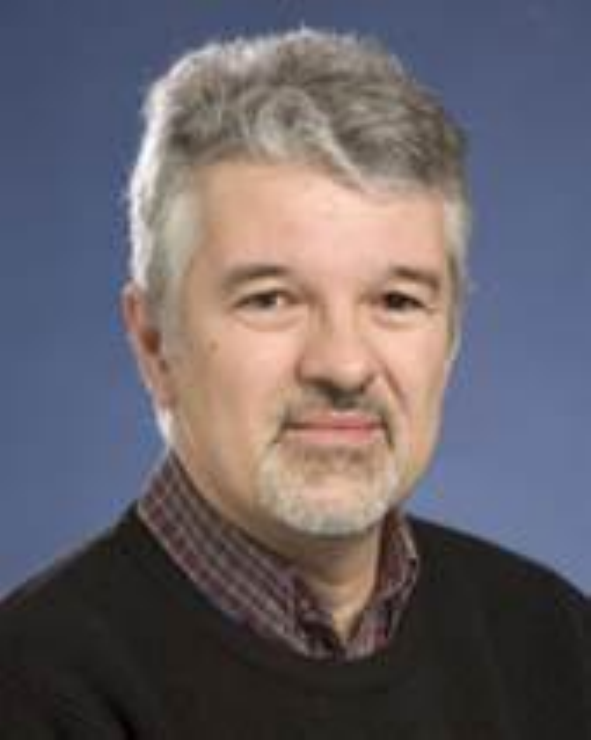}}]{Josef Zuk}
was awarded a B.Sc.(Hons) in Physics and Mathematics from the
University of Melbourne in 1982.
In 1985, he completed a D.Phil.\ in Theoretical Particle Physics at the
University of Oxford, where he remained as a research fellow until 1987,
before moving on to the Max Planck Institute for Nuclear Physics in
Heidelberg, Germany to take up a research fellowship in Theoretical
Condensed-Matter Physics.
In 1990, he commenced a faculty position in the Department of Physics at the
University of Manitoba, Winnipeg, Canada.
Having joined DST Group (then DSTO) in 1995, he is currently a senior research scientist in the
Joint and Operations Analysis Division (JOAD), where he works in operations research,
and is mainly engaged in operational aspects of radar performance modelling and
simulation, particularly for airborne maritime surveillance radar systems.
\end{IEEEbiography}
\vspace{\fill}
\end{document}